\documentclass[fleqn,usenatbib]{mnras}

\usepackage[T1]{fontenc}

\DeclareRobustCommand{\VAN}[3]{#2}
\let\VANthebibliography\thebibliography
\def\thebibliography{\DeclareRobustCommand{\VAN}[3]{##3}\VANthebibliography}

\usepackage{graphicx}	
\usepackage{amsmath}	
\usepackage{amssymb}	

\usepackage{booktabs}

\usepackage{cleveref}
\crefname{section}{§}{§§}
\Crefname{section}{§}{§§}

\usepackage{newtxtext,newtxmath}

\title[LoS effect on the 21-cm signal from radio galaxies]{Strong 21-cm fluctuations and anisotropy due to the line-of-sight effect of radio galaxies at cosmic dawn} 

\author[S. Sikder et al.]{
Sudipta Sikder$^{1}$\thanks{E-mail: sudiptas@mail.tau.ac.il},
Rennan Barkana$^{1,2,3}$,
Anastasia Fialkov$^{4,5}$,
and Itamar Reis$^{1}$
\\
$^{1}$School of Physics and Astronomy, Tel-Aviv University, Tel-Aviv, 69978, Israel \\
$^{2}$Institute for Advanced Study, 1 Einstein Drive, Princeton, New Jersey 08540, USA\\
$^{3}$Department of Astronomy and Astrophysics, University of California, Santa Cruz, CA 95064, USA\\
$^{4}$Institute of Astronomy, University of Cambridge, Madingley Road, Cambridge, CB3 0HA, UK\\
$^{5}$Kavli Institute for Cosmology, Madingley Road, Cambridge CB3 0HA, UK\\}

\date{Accepted XXX. Received YYY; in original form ZZZ}

\pubyear{2022}

\begin{document}
\label{firstpage}
\pagerange{\pageref{firstpage}--\pageref{lastpage}}
\maketitle

\begin{abstract}The reported detection of the global 21-cm signal by the EDGES collaboration is significantly stronger than standard astrophysical predictions. One possible explanation is an early radio excess above the cosmic microwave background. Such a radio background could have been produced by high redshift galaxies, if they were especially efficient in producing low-frequency synchrotron radiation. We have previously studied the effects of such an inhomogeneous radio background on the 21-cm signal; however, we made a simplifying assumption of isotropy of the background seen by each hydrogen cloud. Here we perform a complete calculation that accounts for the fact that the 21-cm absorption occurs along the line of sight, and is therefore sensitive to radio sources lying behind each absorbing cloud. We find that the complete calculation strongly enhances the 21-cm power spectrum during cosmic dawn, by up to two orders of magnitude; on the other hand, the effect on the global 21-cm signal is only at the $5\%$ level. In addition to making the high-redshift 21-cm fluctuations potentially more easily observable, the line of sight radio effect induces a new anisotropy in the 21-cm power spectrum. While these effects are particularly large for the case of an extremely-enhanced radio efficiency, they make it more feasible to detect even a moderately-enhanced radio efficiency in early galaxies. This is especially relevant since the EDGES signal has been contested by the SARAS experiment.
\end{abstract}




\section{Introduction}

The redshifted 21-cm signal which originates due to the hyperfine splitting of the neutral hydrogen in the intergalactic medium (IGM) is the most promising probe of the early universe, most importantly, the epoch of the first stars and the Epoch of Reionization (EoR). The rest-frame frequency of $1420$ MHz is redshifted due to the expansion of the universe and can be detected using ground-based radio telescopes at frequencies below $200$ MHz against the background radiation, which is usually assumed to be the cosmic microwave background (CMB).

The first claimed detection of the all-sky averaged global 21-cm signal from $z \sim 13-17$ was the EDGES low band observation in the frequency range of $50-100$ MHz \citep{bowman18}.  The signal was centered at $z\sim17$ (which corresponds to $\nu\sim78$ MHz) with a strong absorption feature of $T_{21} = -500^{+200}_{-500}$ mK. While disputed at 95\% significance by the SARAS experiment \citep{SARAS3}, with further measurements expected to resolve this tension, the tentative EDGES signal has inspired various theories. Specifically, this anomalously strong trough has two main categories of explanations. One category is that an additional cooling mechanism can cool the gas faster than only adiabatic cooling due to the cosmic expansion. An additional cooling mechanism has been suggested \citep{barkana18, Berlin, barkana18a, munoz18, Liu19, Barkana2022} that involves a non-gravitational interaction between the baryons and dark matter particles (e.g., via Rutherford-like scattering) that drives down the temperature of the gas leading to the strong observed absorption. The other category of explanation is the presence of an excess radio background at high redshifts, well over the CMB level \citep{bowman18, feng18, ewall18, fialkov19, mirocha19, ewall20}. Specifically, \citet{fialkov19} showed that the EDGES signal could be explained by a homogeneous external radio background with a synchrotron spectrum. However, this external radio background is not directly related to astrophysical sources. Exotic processes such as dark matter annihilation or superconducting cosmic strings \citep{Fraser:2018, Pospelov:2018, Brandenberger:2019} could give rise to this kind of homogeneous external radio excess. A more astrophysically-grounded approach is to assume that radio-loud sources such as active galactic nuclei  \citep[AGN,][]{urry95, biermann14, bolgar18, ewall18, ewall20} or star-forming galaxies \citep{condon92, jana19} at high redshift could produce an  excess radio background, which in this case would be inhomogeneous. 
\citet{Reis2020} first incorporated the inhomogeneous excess galactic radio background into semi-numerical simulations of the early Universe, and explored the effect on the global 21-cm signal and on the 21-cm power spectrum. Interestingly, at low frequencies, ARCADE2 \citep{fixsen11, seiffert11} detected an excess radio background over that CMB that was confirmed by LWA1 \citep{dowell18} in the frequency range $40-80$ MHz. This observed excess radio could be explained by extragalactic sources, but it is unclear what fraction of the observed excess originates from Galactic compared to extragalactic sources \citep[e.g.,][]{Subrahmanyan:2013}. 

In our previous work \citep{Reis2020} we made a simplifying approximation and assumed that the effect of the radio background on a given hydrogen cloud can be determined from the isotropically-averaged radio intensity at that position. However, since 21-cm absorption occurs along the line-of-sight (hereafter LoS), the 21-cm effect effectively involves two different radio intensities. The isotropically-averaged radio intensity is appropriate for effects such as the physical heating of the gas, while the calculation of radiative transfer along the line of sight depends on the radio intensity coming only from radio sources lying behind the hydrogen cloud, along our line of sight. This can potentially enhance the 21-cm power spectrum due to the LoS radio fluctuations, especially early in cosmic dawn when the number of contributing radio sources is small. We perform this complete calculation in this work, and also quantify the resulting LoS anisotropy in the 21-cm power spectrum using the anisotropy ratio \citep[following][]{fialkov2015}.

This paper is organized as follows: we briefly describe our semi-numerical simulation in section \ref{sec:method}. In section \ref{sec:21cm_signal}, we review the theoretical framework of the 21-cm signal in the presence of an excess radio background and show how we include the line of sight effect of radio fluctuations in the simulation. In section \ref{sec:result}, we explore the effect of the line of sight radio fluctuations on the 21-cm signal including the anisotropic power spectrum due to this line of sight effect. We conclude the paper with a summary in section \ref{sec:summary}. 

\section{Basic method}\label{sec:method}

We use our semi-numerical 21-cm simulation code \citep[e.g.,][]{visbal12, fialkov14, cohen17, fialkov19} to calculate the 21-cm signal over a wide range of redshifts. This simulation code was originally inspired by {\sc 21cmFAST} \citep{mesinger11}, but it is entirely an independent implementation. The code simulates the realization of the universe in a $384^3$ Mpc$^3$ comoving cosmological volume with a resolution of $3$ comoving Mpc. The simulation is based on the following algorithm: we create a random realization of the large-scale linear density field, i.e., the three dimensional cubes of density fluctuations and the relative velocity between the dark matter and the baryons \citep{tseliakhovich10} given the power spectra of initial Gaussian random density fields and velocity fields \citep[calculated using the publicly available code  {\sc CAMB},][]{camb}. Given the large scale density fields and the relative velocity,  we obtain the population of the collapsed dark matter halos inside each cell of 3$^3$ Mpc$^3$ comoving volume, using the modified Press-Schechter model \citep{press74, sheth99, barkana04}. The baryon fraction contained in each halo is assumed to be the cosmic mean, except that it is reduced due to the streaming velocity \citep{tseliakhovich10,fialkov12}. Star formation takes place where the dark matter halos are massive enough to radiatively cool the infalling gas. This sets the minimum mass of star forming halos (given by a minimum circular velocity $V_c$), except that feedback can also affect this free parameter. Another important parameter is the star formation efficiency, $f_*$, and we also take into account the suppression of star formation due to the above-mentioned relative velocity between dark matter and baryons, Lyman-Werner feedback on molecular-hydrogen cooling halos \citep{haiman97,fialkov2013}, and photoheating feedback \citep{rees86,sobacchi13,cohen16}. 

Once we have a population of galaxies, we calculate the radiation fields emitted by those galaxies. The most relevant radiation fields that affect the 21-cm signal are ionizing, Ly-$\alpha$ and X-ray radiation. To calculate the intensity of the Ly-$\alpha$ radiation field, i.e., $J_{\alpha}$, we assume that galaxies contain population II stars. The X-ray luminosity ($L_{\rm{X}}$) of the galaxies is assumed to scale with the star formation rate (SFR), which is based on X-ray observations of low-redshift galaxies  \citep[e.g.,][]{Grimm, Gilfanov, Mineo:2012, fragos13, fialkov14a, Pacucci:2014}:
\begin{equation}\label{eqn:Lx_SFR_relation}
    \frac{L_{\rm{X}}}{\rm{SFR}} = 3\times10^{40} f_{\rm{X}} \ \rm{erg\ s^{-1} M^{-1}_\odot yr} \ ,
\end{equation}
where the standard normalization factor $f_{\rm{X}}$ is the X-ray efficiency of the sources, a free parameter in our simulation. Here $f_{\rm{X}} = 1$ corresponds to the typical observed value for low metallicity galaxies. In addition to the luminosity, the shape of the X-ray spectral energy density (SED) affects the 21-cm signal. We assume that the shape of the X-ray SED is determined by a power law slope ($\alpha$, which we set equal to 1.5) and a low energy cutoff ($E_{\rm{min}}$). The hard X-ray SED sourced by a population of high redshift X-ray binaries (XRBs) peaks at energy $\sim 3$ keV. Soft X-ray sources (with typical energy $\sim 0.5$ kev) produce strong fluctuations on small scales (up to a few tens of Mpc), whereas the typical mean free path of hard X-ray photons is so large that the fluctuations are reduced and also they lose some of their energy due to the redshift effect. 

After the heating transition due to X-ray photons, the universe starts to reionize. The phase transition known as the epoch of reionization is expected to have occured inside out, meaning that the high-density regions containing most of the sources proceeded to reionize first \citep{barkana04, furlanetto04}. As the 21-cm brightness temperature is proportional to the fraction ($x_{\rm{HI}}$) of the neutral hydrogen atoms in the IGM, the amplitude of the global signal decreases as reionization proceeds. Another free parameter in our simulation is the ionization efficiency, $\zeta$. The late stages of reionization also depend on the maximum mean free path of the ionizing photons, $R_{\rm{mfp}}$ \citep{greig15}. Dense regions of neutral hydrogen (specifically, Lyman-limit systems) that appear at high redshifts due to structure formation, effectively absorb the ionizing radiation and set an upper limit on the effective ionization bubble size. Since here we focus on significantly higher redshifts, we simply set $R_{\rm{mfp}} = 30$ Mpc (comoving) and $\zeta = 30$ for all the cases used in this work; this gives an optical depth to the CMB that is consistent with Planck measurements~\citep{planckcollaboration18}. Finally, an excess radio background above the cosmic microwave background can be included as we discuss in the next section.

\section{21-cm signal}\label{sec:21cm_signal}

The 21-cm brightness temperature, $T_{\rm{21}}$, depends on the contrast between the spin temperature, $T_{\rm{S}}$, of the neutral hydrogen and the background radiation temperature, $T_{\rm{rad}}$, and can be written as 
\begin{equation}
    T_{21} = \frac{T_{\rm S} - T_{\rm rad}}{1+z}(1 - e^{-\tau_{21}})\ .
\label{eq:T21}\end{equation}
Usually the background radiation is assumed to be the CMB (at redshift $z$), in which case $T_{\rm{rad}} = T_{\rm{CMB}} = 2.725(1 + z)$ K, unless there is an excess radio background above the CMB (discussed below). Here $\tau_{21}$ is the optical depth of the 21-cm signal and is given by
\begin{equation}
\tau_{21} = \frac{3h_{\rm pl}A_{10}c \lambda_{21}^2n_{\rm H}}{32\pi k_{B} T_{\rm S} (1+z) dv/dr}\ , \label{eq:tau}
\end{equation}
where $h_{\rm{pl}}$ is the Planck constant, $A_{10}$ is the spontaneous decay rate of the hyperfine transition of the neutral hydrogen, $c$ is the speed of light, $\lambda_{21}$ = $21.1$ cm is the rest frame wavelength of the 21-cm line, $n_{H}$ is the number density of the neutral hydrogen, $k_{B}$ is the Boltzmann constant, $dv/dr = H(z)/(1+z)$ is the gradient of the line of sight component of the comoving velocity field and $H(z)$ is the Hubble constant (Our code also includes fluctuations with respect to this expression for the mean velocity gradient). 

The spin temperature $T_{S}$ can be written as \citep{madau97}
\begin{equation}
T_{S} = \frac{x_{\rm rad} + x_{\rm tot}}{x_{\rm rad} T_{\rm rad}^{-1} + x_{\rm tot} T_{\rm K}^{-1}}\ , \label{eq:TS}
\end{equation}
where 
\begin{equation}
    x_{\rm rad} = \frac{1 - e^{-\tau_{21}}}{\tau_{21}}
\end{equation}
is the radiative coupling \citep{venumadhav18}, and the coupling coefficient $x_{\rm{tot}}$ is the sum of the contributions of Ly-$\alpha$ coupling ($x_{\alpha}$) and the collisional coupling ($x_{c}$), i.e., $x_{\rm{tot}} = x_{\alpha} + x_{c}$, with 
\begin{equation}
    x_{\alpha} = \frac{1}{A_{10} T_{\rm rad}} \frac{16 \pi^2 T_{*} e^2 f_{\alpha}}{27 m_{e} c}  J_{\alpha}\ , \label{eq:xa}
\end{equation}
and 
\begin{equation}
\label{eq:xc}
    x_c = \frac{1}{A_{10} T_{\rm rad}} \kappa_{1-0}(T_{\rm K}) n_{\rm H} T_{\star}\ .
\end{equation}
Here $f_{\alpha} = 0.4162$ is the oscillator strength of the Ly-$\alpha$ transition, $J_{\alpha}$ is the intensity of the Ly-$\alpha$ radiation, $T_* = 0.0682$K and $\kappa_{1-0}(T_K)$ is the known atomic coefficient \citep{allison69, zygelman05}.

When the optical depth $\tau_{21} << 1 $, the 21-cm brightness temperature is given by
\begin{multline}
T_{\rm 21}  \approx  26.8 \left(\frac{\Omega_{\rm b} h}{0.0327}\right)
\left(\frac{\Omega_{\rm m}}{0.307}\right)^{-1/2}
\left(\frac{1 + z}{10}\right)^{1/2} \\
(1 + \delta) x_{\rm HI} \frac{x_{\rm tot}}{1 + x_{\rm tot}} \left( 1 - \frac{T_{\rm rad}}{T_{\rm K}}\right) ~{\rm mK}\ , \label{eq:linear}
\end{multline}
where we have included the effect of the neutral hydrogen fraction $x_{\rm{HI}}$, and of the density contrast $\delta$.

When calculating the kinetic gas temperature we include the usual effects of adiabatic evolution, Compton heating, and X-ray heating. The effect of the radio background on the kinetic gas temperature based on the CMB heating mechanism introduced by \citet{venumadhav18} is also included here \citep[but see objections by][]{Meiksin21}. The heating rate due to the radio background is 
\begin{equation}
    \epsilon_{\rm rad} = \frac{x_{\rm HI} A_{10} }{2 H(z)} x_{\rm rad} \left(\frac{T_{\rm rad}}{T_{\rm S}} - 1 \right) \frac{T_{21}}{T_{\rm K}}\ . \label{eq:heat}
\end{equation}
In practice we use eq.~\ref{eq:tau}, including the effect of the inhomogeneous density and velocity gradient, and do not assume the linearized form as in eq.~\ref{eq:linear}; we note, though, that the linearized expression is in most cases rather accurate.

\subsection{The excess radio background: previous work}

In the presence of an excess radio background, we can rewrite the background radiation temperature $T_{\rm{rad}}$, as 
\begin{equation}
T_{\rm{rad}} = T_{\rm{Radio}} + T_{\rm{CMB}} \ , 
\label{Eq:rad}
\end{equation}
where $T_{\rm{Radio}}$ is the brightness temperature of the excess radio background. In our previous work we calculated the isotropically-averaged radio intensity at each pixel, and used the resulting $T_{\rm{Radio}}$ in all of the above equations. 

One type of excess radio background that has been considered is a homogeneous external radio model that is not directly related to astrophysical sources. This excess radio background could possibly be generated by exotic processes, e.g., annihilating dark matter or super-conducting cosmic strings \citep{Fraser:2018, Pospelov:2018, Brandenberger:2019}. A simple formulation of such a model \citep{fialkov19} sets the brightness temperature of the excess radio background at the 21-cm rest frame frequency at redshift $z$ as 
\begin{equation}
T_{\rm Radio} = 2.725 (1+z)\, A_{\rm r}\, \times \left[\frac{1420}{78 (1+z)}\right]^{\beta} ~{\rm K}\ ,
\label{Eq:ar}
\end{equation}
where $2.725$ K is the CMB temperature today, the spectral index of the synchrotron spectrum is $\beta = -2.6$ (set to match the slope of the observed extragalactic radio background, so that the exotic excess background is consistent with observational limits), and $A_r$ measures the amplitude of the radio background (relative to the CMB at the central redshift of the EDGES claimed absorption feature). 

An excess radio background over the CMB can also be produced by high redshift galaxies if they emit strongly in the radio \citep{Reis2020}. Based on the empirical relation of \citet{gurkan18}, we can write the galaxy radio luminosity per unit frequency, which is proportional to the star formation rate (SFR), as
\begin{equation}
    L_{\rm Radio} (\nu, z ) = f_{\rm Radio} 10^{22} \left(\frac{\nu}{150\, {\rm MHz}}\right)^{-\alpha_{\rm Radio}} \frac{\rm SFR}{M_{\odot}\, \rm{yr}^{-1} }\ ,
    \label{eq:fRadio}
\end{equation}
in units of W Hz$^{-1}$.  In eq.~\ref{eq:fRadio}, the spectral index in the radio band $\alpha_{\rm{Radio}}$ is set to $0.7$ as in \citet{mirocha19} and \citet{gurkan18}; see also \citet{condon02} and \citet{heesen14}. Here $f_{\rm{Radio}}$ is the normalization of the radio emissivity, where $f_{\rm{Radio}} = 1$ for present-day star-forming galaxies. In our work, we assume for simplicity a uniform value of $f_{\rm{Radio}}$, though we note that there is significant scatter in $f_{\rm{Radio}}$ from observations. 

In our previous work \citep{Reis2020}, the brightness temperature of the radio background at redshift $z$ at the 21-cm frequency was calculated by summing the contribution from all the galaxies within the past light-cone \citep[following][]{ewall20}:
\begin{multline}
\label{eq:Tgal}
    T_{\rm Radio} (\nu_{21}, z) = \frac{\lambda_{21}^2}{2 k_{\rm B}} \frac{c (1+z)^3}{4 \pi} \\ \int\epsilon_{\rm Radio}\left(\nu_{21} \frac{1 + z_{\rm em}}{1 + z}, z_{\rm em}\right) (1 + z_{\rm em})^{-1} H(z_{\rm em})^{-1} dz_{\rm em}\ ,
\end{multline}
where $z_{\rm em} > z$ is the redshift at which a photon was emitted, and $\epsilon_{\rm Radio}$ is the comoving radio emissivity, i.e., the luminosity per unit frequency per unit comoving volume, averaged over radial shells within this spherical integral. The radius of each spherical shell is given by the light travel distance between $z_{\rm em}$ and $z$. This calculation is thus similar to that for finding the Ly-$\alpha$ and X-ray radiation fields in our semi-numerical simulation, except that for the Ly-$\alpha$ radiation field, modified window functions are used in order to include the effect of multiple scattering of the Ly-$\alpha$ photons \citep{reis20b}.

\subsection{LoS effect of the radio background from galaxies}

As outlined in the previous subsection, in our previous work we calculated the isotropically-averaged radio intensity at each pixel, and used the resulting $T_{\rm{Radio}}$
in eq.~\ref{Eq:rad}. This is accurate for all the direct physical effects of the radiation, i.e., in eqs.~\ref{eq:xa}, \ref{eq:xc}, \ref{eq:TS}, and \ref{eq:heat}. However, it is only approximately true in the radiative transfer equation~\ref{eq:T21}. 

As the 21-cm absorption occurs along the line of sight, in this work we accurately consider the line of sight contribution of the excess radio background originating from high redshift galaxies, and examine the effect on the 21-cm signal. 
Our previous approximation becomes accurate in the limit of many radio sources reaching each 
pixel, since in that case the radio background does become nearly isotropic. However, at the highest redshifts the radio background at a point is still dominated by a small number of nearby sources, and it then matters whether a source is behind our line of sight to the absorbing pixel, or not. Indeed, we expect the LoS effect to amplify the 21-cm fluctuations at early times.

Here we will continue to refer to the isotropically-averaged radio intensity at a given pixel as $T_{\rm{Radio}}$, but we also use a different window function and calculate the brightness temperature of the radio background from sources lying behind the pixel along our LoS; we refer to the latter quantity as  $T_{\rm{R,los}}$. For clarity, we first find the observed 21-cm brightness temperature relative to the CMB:
\begin{equation}
T_{21}^{\rm CMB} = \frac {\left( T_{R,\rm los} + T_{\rm CMB} \right) e^{- \tau_{21}} + T_S \left( 1 - e^{- \tau_{21}} \right) - T_{\rm CMB}} {1+z}\ ,
\end{equation}
where here $T_S$ and $\tau_{21}$ depend on $T_{\rm{Radio}}$. Now, if there is indeed a strong excess radio background, it too is observed and is subtracted out in any method of foreground removal (since it is assumed here to have a smooth power-law synchrotron spectrum). Another way of expressing this is that we must subtract out the $\tau_{21}=0$ case in order to arrive at the final expression:
\begin{equation}
T_{21} = \frac { 
 T_S - \left( T_{R,\rm los} + T_{\rm CMB} \right) } {1+z}\ \left( 1 - e^{- \tau_{21}} \right)\ .
 \label{eq:T21new}
\end{equation}
This replaces eq.~\ref{eq:T21} (which also was written after subtracting out the $\tau_{21}=0$ case under the previous approximation). Since the 21-cm optical depth is usually quite small, we also note the linearized form of this full expression including the line-of-sight effect, which replaces eq.~\ref{eq:linear}:
\begin{multline}
T_{\rm 21}  \approx  26.8 \left(\frac{\Omega_{\rm b} h}{0.0327}\right)
\left(\frac{\Omega_{\rm m}}{0.307}\right)^{-1/2}
\left(\frac{1 + z}{10}\right)^{1/2} (1 + \delta) x_{\rm HI} \frac{1}{1 + x_{\rm tot}} \\
\left [ x_{\rm tot} \left( 1 - \frac{T_{R,\rm los} + T_{\rm CMB}}{T_{\rm K}}\right) + \left( 1 - \frac{T_{R,\rm los} + T_{\rm CMB}}{T_{\rm{Radio}} + T_{\rm{CMB}}}\right) \right ]~{\rm mK}\ . \label{eq:linearNew}
\end{multline}
As we noted above, the last two equations also contain implicit dependencies on $T_{\rm{Radio}}$ (the isotropically-averaged radio intensity) through eq.~\ref{Eq:rad}, which affects eqs.~\ref{eq:xa}, \ref{eq:xc}, \ref{eq:TS}, and \ref{eq:heat}. 

The last factor in eq.~\ref{eq:linearNew} shows how the spin temperature varies between $T_{\rm{Radio}}+ T_{\rm CMB}$ (in the uncoupled case) and $T_{\rm K}$ (after saturated coupling), while the observations always probe the contrast between $T_S$ and $T_{R,\rm los}+ T_{\rm CMB}$. The uncoupled case (where the previous result gave no signal) shows explicitly how the line-of-sight effect provides a new source of 21-cm fluctuations. Meanwhile, the last factor in eq.~\ref{eq:linearNew} together with eqs.~\ref{eq:xa} and \ref{eq:xc} shows that the Ly-$\alpha$ coupling transition (when $x_{\rm tot} \sim 1$) is substantially delayed by an intense radio background. Also, the sign transition (when the mean 21-cm signal goes from absorption to emission) is significantly delayed, as it no longer occurs around the heating transition (normally defined
as the average gas temperature $T_{\rm K}$ reaching $T_{\rm CMB}$), but must wait for the gas to heat to the higher temperature
given by the average of $T_{R,\rm los} + T_{\rm CMB}$ (assuming $x_{\rm tot}$ is large at that time). Even if CMB heating is effective (given the radio-background boost in eq.~\ref{eq:heat}), the sign transition is still substantially delayed when the excess radio background is strong.  

\subsection{LoS anisotropies in the 21-cm signal}

Among the various sources of fluctuations that contribute to the 21-cm fluctuations, many are statistically isotropic. This is true, for example, for the gas density fluctuations that arise from the initial conditions and are a potential probe of the cosmological parameters. Once the first stars and galaxies form, various forms of radiation including Ly-$\alpha$ photons, ionizing photons, and X-ray photons, become additional sources of 21-cm fluctuations. These radiation fields are produced by the processes of star and galaxy formation that are complex and non-linear, but have no overall preferred directions. Thus, the 21-cm signal due to these radiation fluctuations is isotropic. However, since the 21-cm signal is redshifted and is determined by LoS absorption, there are a number of effects that make it anisotropic. It is important to quantify the LoS anisotropy in the redshifted 21-cm signal for a better understanding of the 21-cm signal itself and also since the anisotropy is potentially directly observable with upcoming telescope arrays such as the Square Kilometre Array (SKA). 

The coherent inflow of matter into over-dense regions and outflow of matter from under-dense regions, namely the peculiar velocity of the baryonic matter, makes the 21-cm signal anisotropic along the LoS. As a result, the 21-cm power spectrum is expected to be anisotropic due to the radial component of the peculiar velocity gradient \citep{SB_SSA_2004, barkana2005}. The light-cone effect, whereby only the LoS direction corresponds to a varying redshift, also produces a LoS anisotropy in the 21-cm fluctuations \citep{barkana_loeb2006, Datta12}. Due to the uncertainty in the values of the cosmological parameters, another potentially observable source of anisotropy in the 21-cm power spectrum is the Alcock-Paczy\'{n}ski effect \citep{alcock_paczynski,Nusser,APindian,MeAP}. In our calculations we include the main expected source of 21-cm anisotropy out of these, i.e., the LoS anisotropy due to peculiar velocities. Now, since we consider in this work the line of sight dependence of the radio fluctuations, this naturally introduces a new potential sources of anisotropy in the 21-cm signal. 

\begin{figure*}
    \centering
 \includegraphics[width=0.96\textwidth]{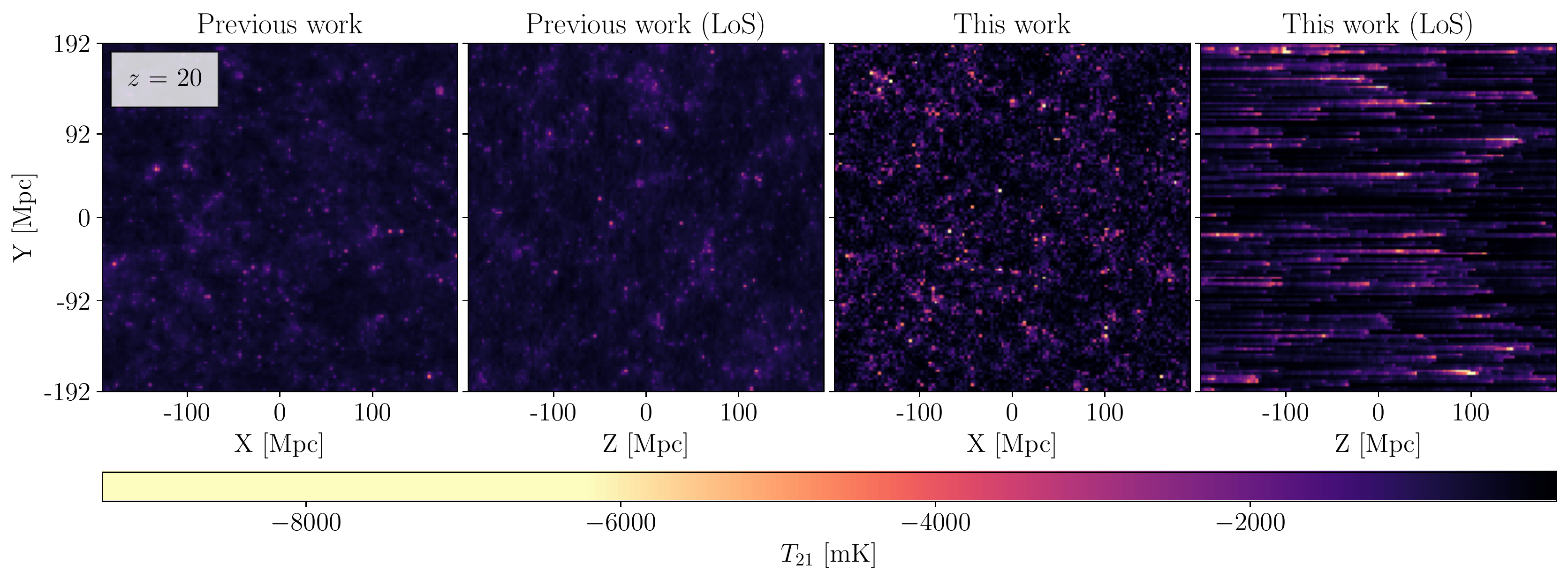}
 
    \caption{Line-of-sight effect of fluctuations in the excess radio background on the cosmic dawn 21-cm signal, illustrated at $z=20$. We compare the case of a fluctuating radio background from our previous work ({\bf left two panels}) to the full calculation including the LoS effect of the fluctuating radio background emitted by galaxies ({\bf right two panels}). From a single cubic simulation box, we show both the 21-cm slice perpendicular to the LoS ($XY$ plane) and one that includes the LoS ($Z$) direction. Both models have the same radio production efficiency, with astrophysical model parameters:  $V_c = 16.5$ km s$^{-1}$, $f_* = 0.1$, and $f_{\rm Radio} = 3000$; note that the circular velocity corresponds to the minimum halo mass for star formation being set by atomic cooling, and at $z = 20$ the mass is $3\times10^7 M_{\odot}$. In this example we show the 21-cm slices from $z = 20$, when X-ray heating and reionization are rather insignificant, but for completeness we note that the parameters are hard X-rays ($E_{\rm{min}} = 1$ keV with $f_{\rm{X}} = 1$) and reionization parameters as noted in section~\ref{sec:method}. We also note that we show the $50$'th slice from the simulation box along each axis, i.e., centered at $Z=-43.5~$Mpc (for the $XY$ plane) and $X=-43.5~$Mpc (for the $ZY$ plane). 
}
    \label{fig:21cm_slice_hard}
\end{figure*}

In order to understand the anisotropy, we first note that in the presence of the anisotropy due to peculiar velocities, the 21-cm power spectrum (in linear theory) can be written as a polynomial $P(k, \mu)$, where $\mu$ is the cosine of the angle between $\mathbf{k}$ and the line of sight \citep{barkana2005}:
\begin{equation}
P(k, \mu) = P_{\mu^0}(k) + \mu^2 P_{\mu^2}(k) + \mu^4 P_{\mu^4}(k) \ .
\end{equation}
Here $P_{\mu^0}(k)$ results from the fluctuations from all the isotropic sources, $P_{\mu^4}(k)$ is proportional to the primordial density power spectrum, and $P_{\mu^2}(k)$ is proportional to the correlation between the density and the radiation from the isotropic sources. This decomposition of the 21-cm power spectrum can potentially provide valuable information beyond just the spherically-averaged power spectrum \citep{barkana2005, fialkov2015}. However, in the presence of non-linearities and more generally, \citet{fialkov2015} proposed a simpler method to measure the anisotropy in the 21-cm power spectrum by defining the anisotropy ratio: 
\begin{equation}\label{eqn:anisotropic_ratio}
    r_{\mu} (k, z) \equiv \frac{\langle P(\mathbf{k}, z)_{| \mu_k|>0.5}\rangle}{\langle P(\mathbf{k}, z)_{|\mu_k|<0.5}\rangle} - 1 \ .
\end{equation}
Here the angular brackets denote an angular averaging over a range of $\mu$ values. The value of $r_{\mu}(k, z)$ captures in one number (at each $k$ and $z$) the overall angular dependence of the power spectrum. If $r_{\mu}$ is close to zero (i.e., much smaller than unity in absolute value), the power spectrum shows little angular dependence, when it is large and positive the fluctuations are stronger along the LoS, and when it is large and negative the fluctuations are stronger in directions on the sky (i.e., perpendicular to the LoS). In this work, we use equation \ref{eqn:anisotropic_ratio} in order to quantitatively explore the anisotropy in the 21-cm power spectrum. A more detailed analysis that quantifies the anisotropy using other methods is left for future work.

\section{Results}\label{sec:result}

\subsection{Implications of the LoS effect of the radio background for the 21-cm signal}

In order to study the impact of an excess radio background from early galaxies on the 21-cm signal, we compare several simulated cases. We start with the cases considered in \citet{Reis2020}, i.e., the CMB-only case (without any radio excess) and the radio excess case in the isotropically-averaged approximation of eq.~\ref{eq:T21}, and compare them to the fully accurate calculation including the LoS effect as in eq.~\ref{eq:T21new}. We find that the LoS effect on radio fluctuations can significantly affect the 21-cm signal in the redshift range relevant for current and upcoming radio telescopes. However, this effect varies between astrophysical models, scales and different epochs. Here we mainly focus on the dependence on the radio efficiency parameter $f_{\rm Radio}$ from eq.~\ref{eq:fRadio}. In \citet{Reis2020} we found that models that can explain the EDGES low-band absorption require $f_{\rm Radio} \times f_* \sim 140$ or higher (depending on the other astrophysical parameters). As our main case we consider $f_{\rm Radio}=3000$ (with $f_*=0.1$), which lies well within the range compatible with EDGES. We also, though, consider much lower $f_{\rm Radio}$ values, that are still well above unity (i.e., moderately enhanced compared to low-redshift galaxies) but do not depend on the veracity of the EDGES measurement. We emphasize that there are few observational constraints on the radio efficiency of galaxies at very high redshifts, and given the very different astrophysical conditions at that epoch compared to those at low redshift, it is important to keep an open mind on the possible radio efficiency, until new observational constraints can be established. 
    
Fig. \ref{fig:21cm_slice_hard} shows a comparison of 21-cm slices at $z=20$ for $f_{\rm Radio}=3000$, simulated using two different models of radio fluctuations: as in our previous work \citep[][two left-most panels]{Reis2020}, or as in this work including the line-of-sight effect (two right-most panels). In this case in which early galaxies were unusually bright in low-frequency radio emission relative to star formation, the resulting radio background strongly enhances the 21-cm signal map, and brings out the regions surrounding early radio galaxies as strong peaks of 21-cm absorption. The full inclusion of the LoS effect further brightens these regions by a factor of 2 or 3 in 21-cm brightness temperature. Especially interesting is the clear induced structure along the line of sight, which is potentially a clear observational signature of the presence of a strong background from high-redshift radio galaxies. However, these slices represent pure theoretical predictions of the 21-cm signal, and do not include observational effects that are expected to make this structure significantly more difficult to discern, as we explore further below.

At high redshifts during the epoch of the first stars, the 21-cm signal is normally dominated by Ly-$\alpha$ fluctuations along with some contributions from the density and temperature fluctuations. As shown in the previous section, in the presence of a strong radio background, the LoS effect produces 21-cm fluctuations even before significant Ly-$\alpha$ coupling, and this then mixes in with  Ly-$\alpha$ fluctuations (in an interplay of the two terms in the
last factor in eq.~\ref{eq:linearNew}). At this stage we find that the 21-cm signal is enhanced by up to a factor of a few by the radio background; once the delayed Ly-$\alpha$ coupling does occur (i.e., $x_{\rm tot} \sim 1$ is reached) in some region, the enhancement can become much larger (in proportion to the radio intensity), but this is tempered by the stronger CMB-radio heating. We note that the radio and Ly-$\alpha$ intensity fluctuations are positively correlated (also with the underlying density fluctuations), since both fields originate from the same high redshift galaxies. Thus, these fluctuations enhance each other. However, the 
temperature fluctuations (whether from CMB-radio heating or, later on, from X-rays) work mostly in the opposite direction due to the inverse gas temperature dependence of the 21-cm signal (as long as it is an absorption signal, relative to the radio background). When we consider the line-of-sight effect of the radio background (which is unique and does not occur in the other relevant radiation backgrounds), it is positively correlated with the radio intensity (since a nearby source along the line of sight contributes also to the overall radio intensity), but it adds a strong source of random fluctuations due to the directional dependence.

\begin{figure}
    \centering
    \includegraphics[width=0.48\textwidth]{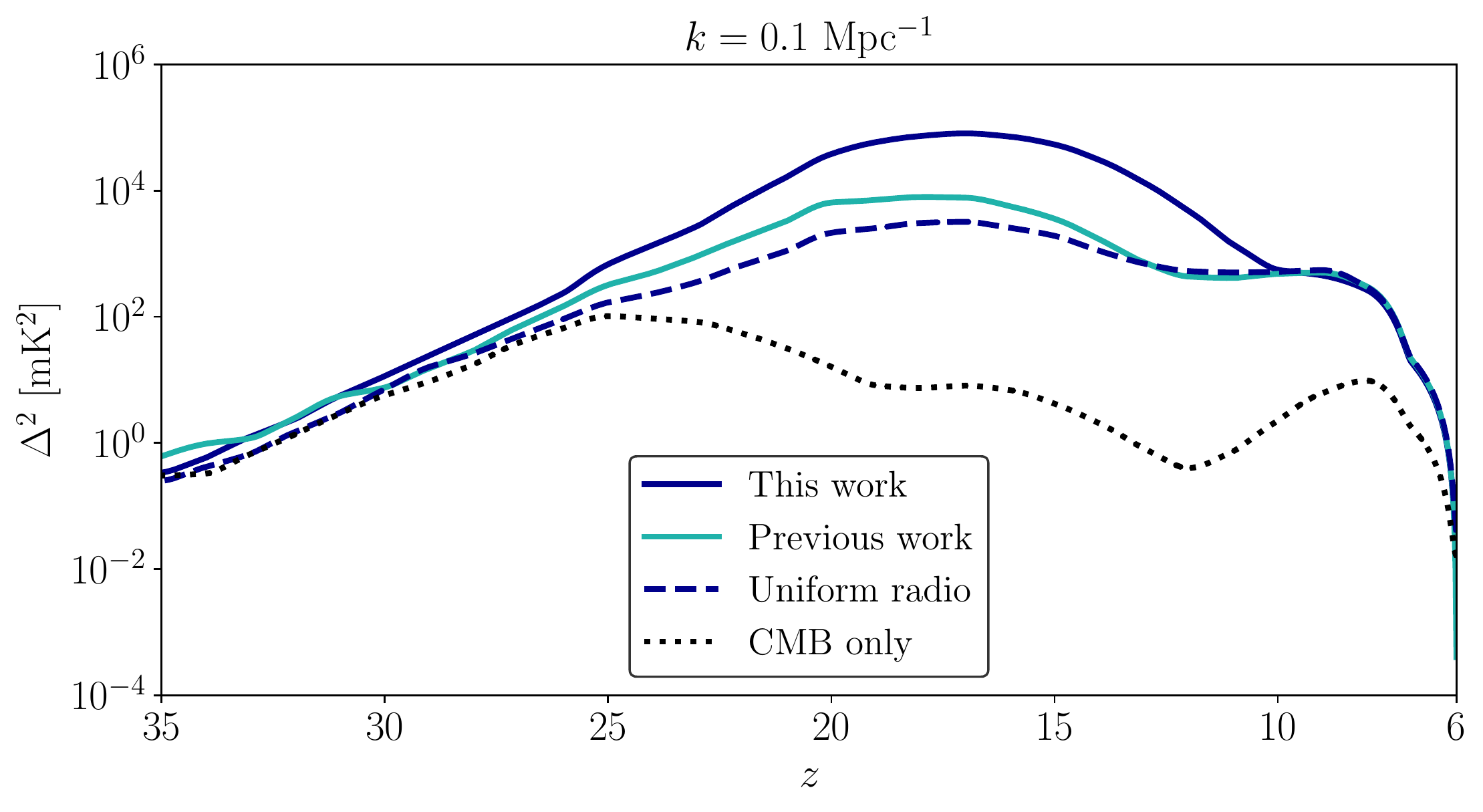}
    \includegraphics[width=0.48\textwidth]{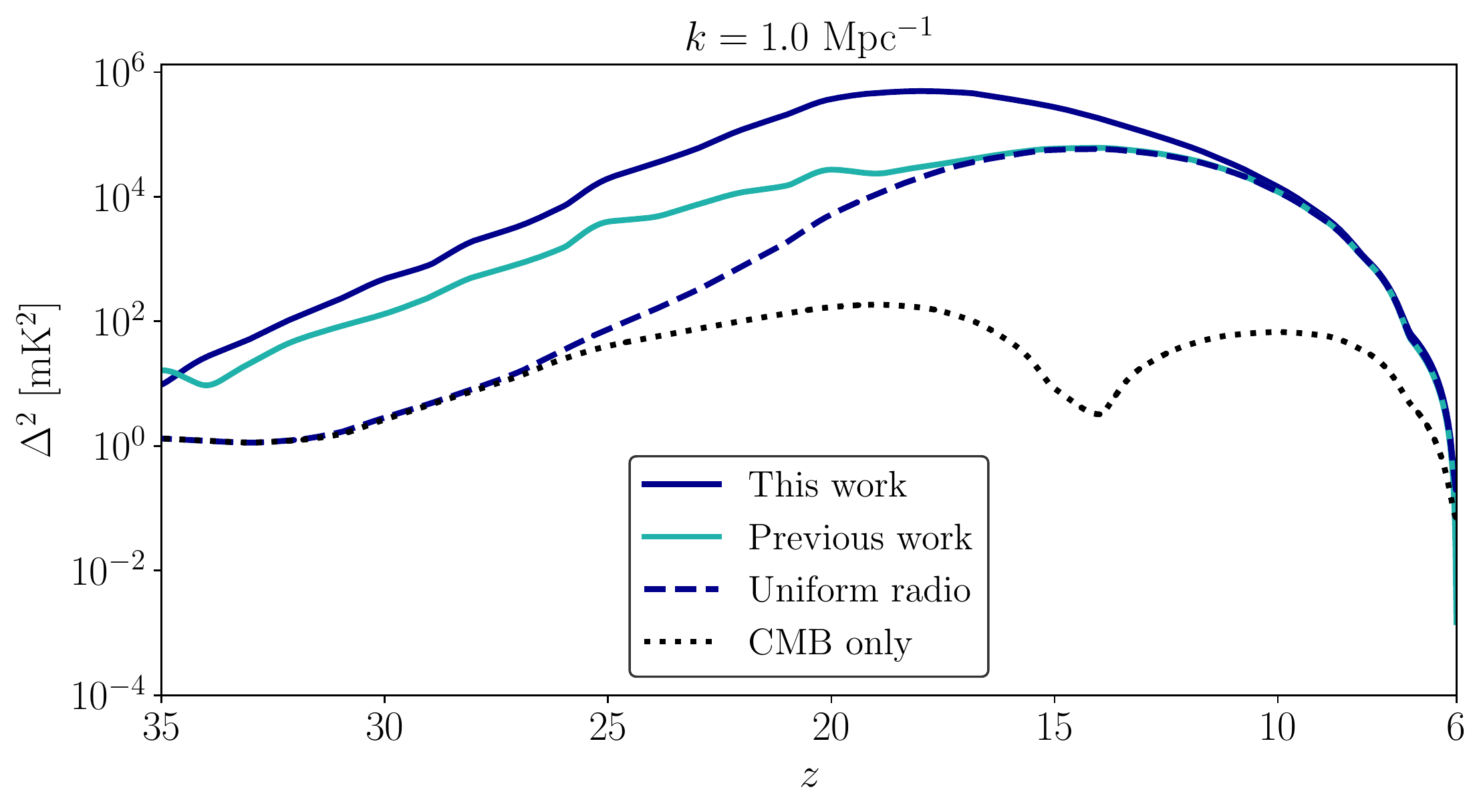}
    \caption{The 21-cm power spectrum at two wavenumbers, $k=0.1 \ \rm{Mpc^{-1}}$ and $k=1.0 \ \rm{Mpc^{-1}}$, as a function of redshift, for various simulation runs with a fixed excess radio background of $f_{\rm{Radio}} = 3000$. We compare the fluctuating radio background with the LoS effect (solid blue line) to the fluctuating radio background from our previous work (solid cyan line). For added comparison, the 21-cm power spectrum due to a uniform excess radio background is shown (dashed blue line), along with the standard case with no excess radio background, i.e., the CMB only case (dotted black line). The uniform radio background case has the same mean intensity of the excess radio at each redshift as in the cases of the fluctuating radio background. Shown here is the model with a hard X-ray SED with $E_{\rm{min}} = 1$ keV.}
    \label{fig:power_spectrum_z}
\end{figure}

The statistics of the fluctuations are well captured in the 21-cm power spectrum, shown in Fig.~\ref{fig:power_spectrum_z} for two different values of wavenumber, as a function of redshift. In addition to showing the enhanced radio background from this work (with the LoS effect) compared to our previous work, here we also show two other comparison cases. One is the case without an excess radio background, i.e., the CMB only case with $f_{\rm{Radio}} = 0$ (dotted black line in both panels). The other comparison case is the uniform radio case, i.e., when we take the cosmic mean excess radio background (which is the same for this work and our previous work) and spread it out evenly over the entire box. This allows us to separate the effect of the overall enhanced radiative background from the effect of the fluctuations in this background. We find that the radio fluctuations boost the 21-cm power spectrum at high redshifts, from the beginning of cosmic dawn on small scales ($k=1$ Mpc$^{-1}$) and from $z \sim 25$ even on large scales ($k=0.1$ Mpc$^{-1}$). At $k=0.1$ Mpc$^{-1}$, the enhancement during cosmic dawn would be by 2 orders of magnitude even with a spatially uniform excess radio background, but the radio fluctuations add more than an additional order of magnitude, mostly due to the line of sight effect of the radio fluctuations. At $k=1$ Mpc$^{-1}$ the enhancement due to radio fluctuations starts earlier, but the behaviour is quite similar at $z=10-20$. Regardless of scale, the radio fluctuations eventually die down as the number of sources becomes large, and the power spectrum becomes the same as it would be with a uniform excess radio background.

\begin{figure}
\centering
    \includegraphics[width=0.48\textwidth]{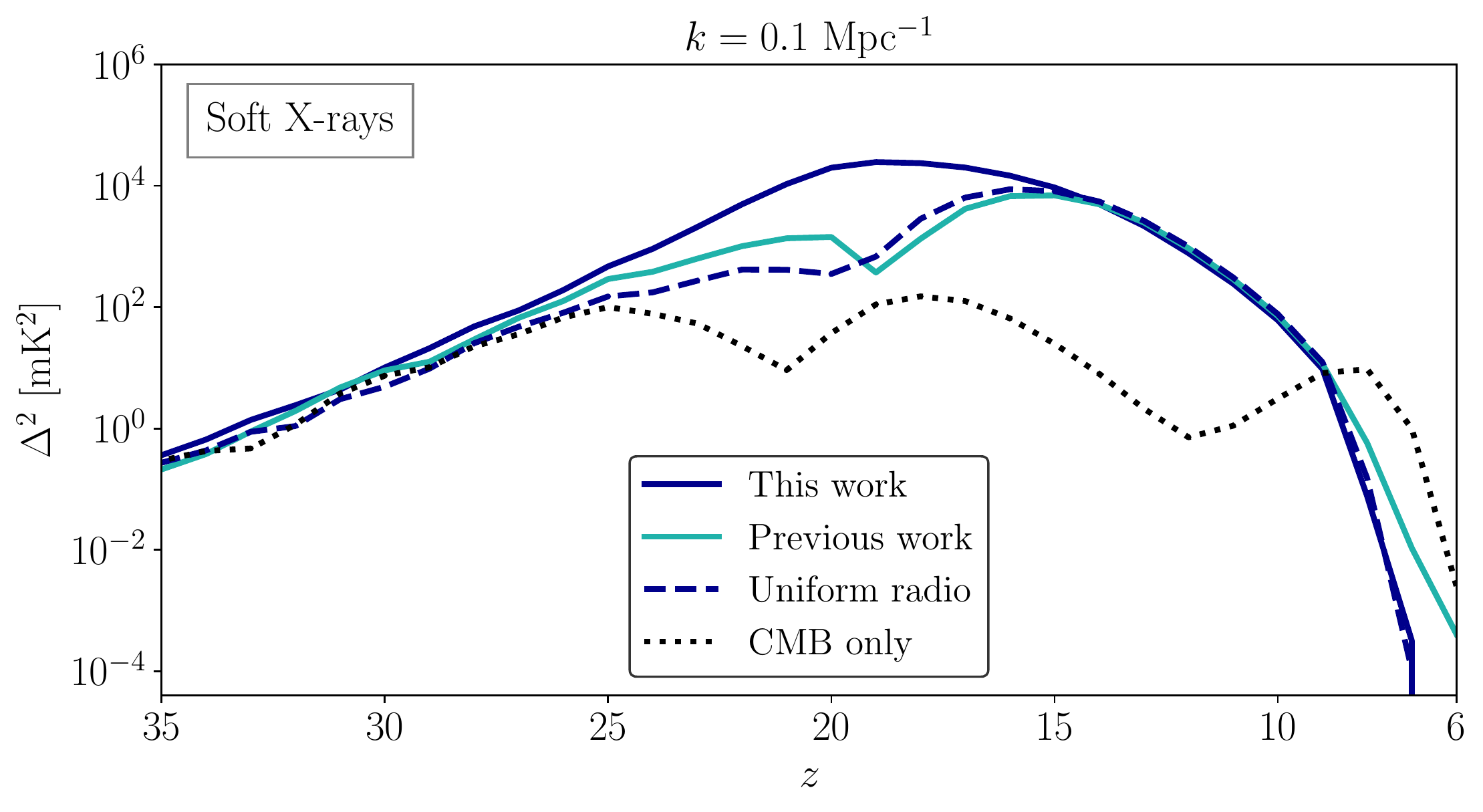}
    \includegraphics[width=0.48\textwidth]{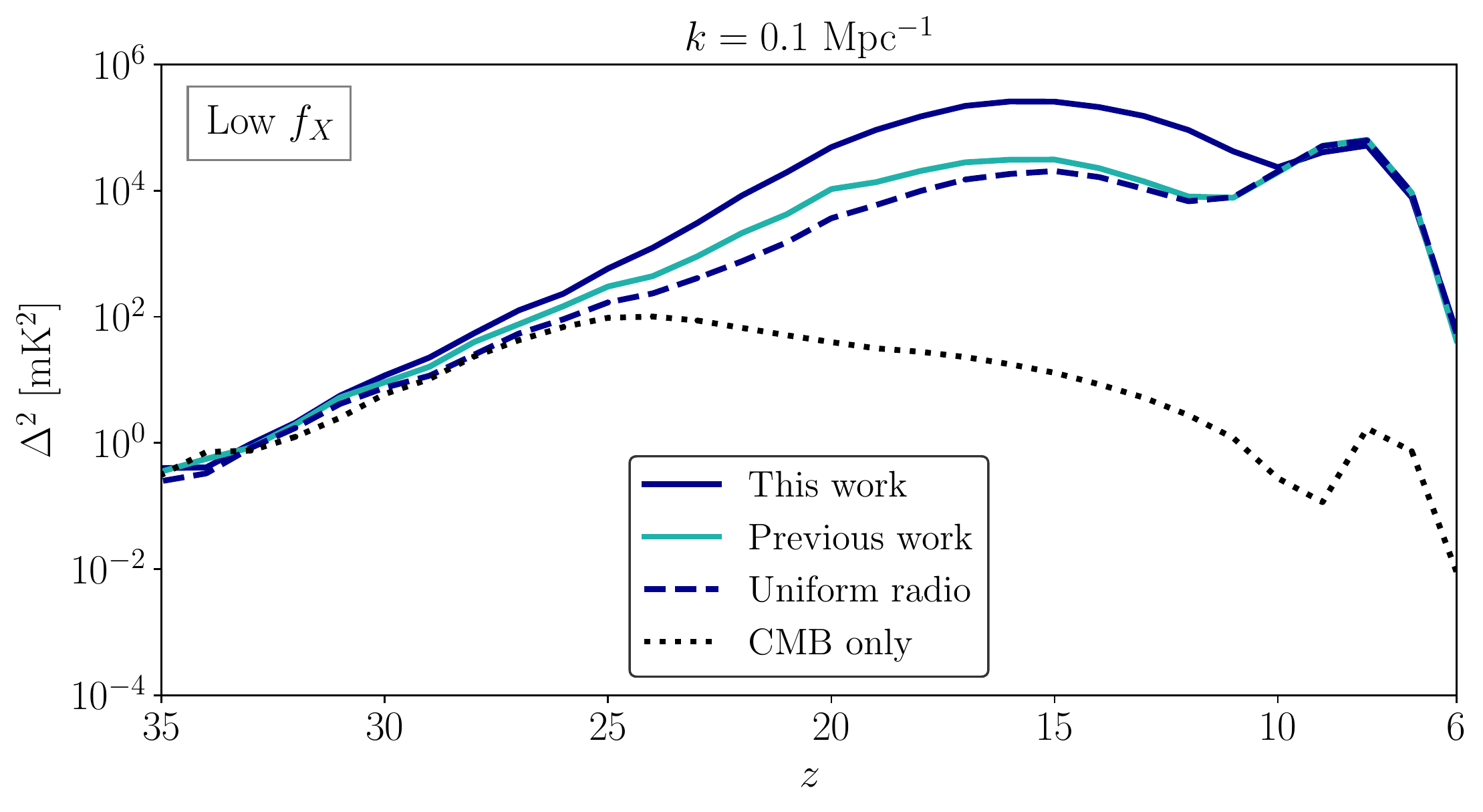}
    \caption{The 21-cm power spectrum at $k = 0.1$ Mpc$^{-1}$ as a function of redshift for various X-ray parameters. \textbf{Top panel:} a soft X-ray SED ($E_{\rm{min}} = 0.1$ keV) with the fiducial X-ray efficiency $f_{\rm{X}} = 1$. \textbf{Bottom panel:} a hard X-ray SED ($E_{\rm{min}} = 1.0$ keV) with a low $f_{\rm{X}} = 0.01$.}\label{fig:power_spectrum_z_fx}
\end{figure}

In order to explore some of the dependence on the various unknown astrophysical parameters, we show the case of a soft X-ray SED in the top panel of Fig. \ref{fig:power_spectrum_z_fx}, for $k=0.1$ Mpc$^{-1}$. As noted above, the heating fluctuations are anti-correlated with the other 21-cm fluctuations, so once the first generation of X-ray sources heat up the IGM, this heating mechanism reduces the 21-cm fluctuations, and then produces a 21-cm fluctuation peak when the heating fluctuations dominate. X-ray photons with lower energies are absorbed locally, while hard X-ray photons ($\gtrsim 1$ keV) lose their energy due to redshifting as they have a much longer mean free path. As a result, heating is delayed and the resulting fluctuations are smaller for a hard X-ray SED compared to a soft X-ray SED. Thus, the CMB-only case does show three peaks (Ly-$\alpha$ coupling, heating, and reionization), but the strong heating peak in the soft X-ray case is barely present in the case of hard X-rays.

In the presence of a strong radio background, the coupling transition is delayed due to the inverse dependence of the coupling coefficients on the radiation background, and the heating transition is delayed as well since the kinetic temperature now needs to reach the higher value $ T_{\rm{CMB}} + T_{\rm{radio}}$ (where $T_{\rm{radio}}$ is replaced by $T_{\rm{R,los}}$ in the full calculation with the LoS effect). At the same time, the fluctuations in the radio background compete with, and sometimes dominate over, the Ly-$\alpha$ and heating fluctuations. For hard X-rays, the two normal peaks are washed out and a single overall peak appears, close to the redshift of the CMB-only heating peak. When we include the LoS radio fluctuations, this overall peak gets a significant boost (and a slight delay). In the case of a soft X-ray SED, without the LoS effect (i.e., in the previous work case) the strong heating fluctuations dominate and maintain a clear heating peak in the 21-cm power spectrum, which is boosted and delayed compared to the case without excess radio (CMB only). However, when we take into account the LoS radio fluctuations, these fluctuations become strong enough compared to the heating fluctuations to again wash out the heating peak from the power spectrum and produce a single overall peak.

Besides the X-ray SED parameters, the X-ray radiation efficiency ($f_{\rm X}$) is another free parameter in our simulation, defined in equation \ref{eqn:Lx_SFR_relation}. In the bottom panel of Fig. \ref{fig:power_spectrum_z_fx} we consider a low efficiency case ($f_{\rm X} = 0.01$, compared to our fiducial $f_{\rm X} = 1$). In this case the heating peak disappears from the 21-cm power spectrum even in the CMB only case. In the presence of a strong radio background, the dominance of the radio fluctuations (in this case with hard X-rays) means that lowering the X-ray efficiency has little effect at high redshifts, but it does boost the low-redshift signal since in this case the 21-cm signal is maintained in absorption down to the reionization epoch. 

\begin{figure*}
    \centering
    \includegraphics[width=0.48\textwidth]{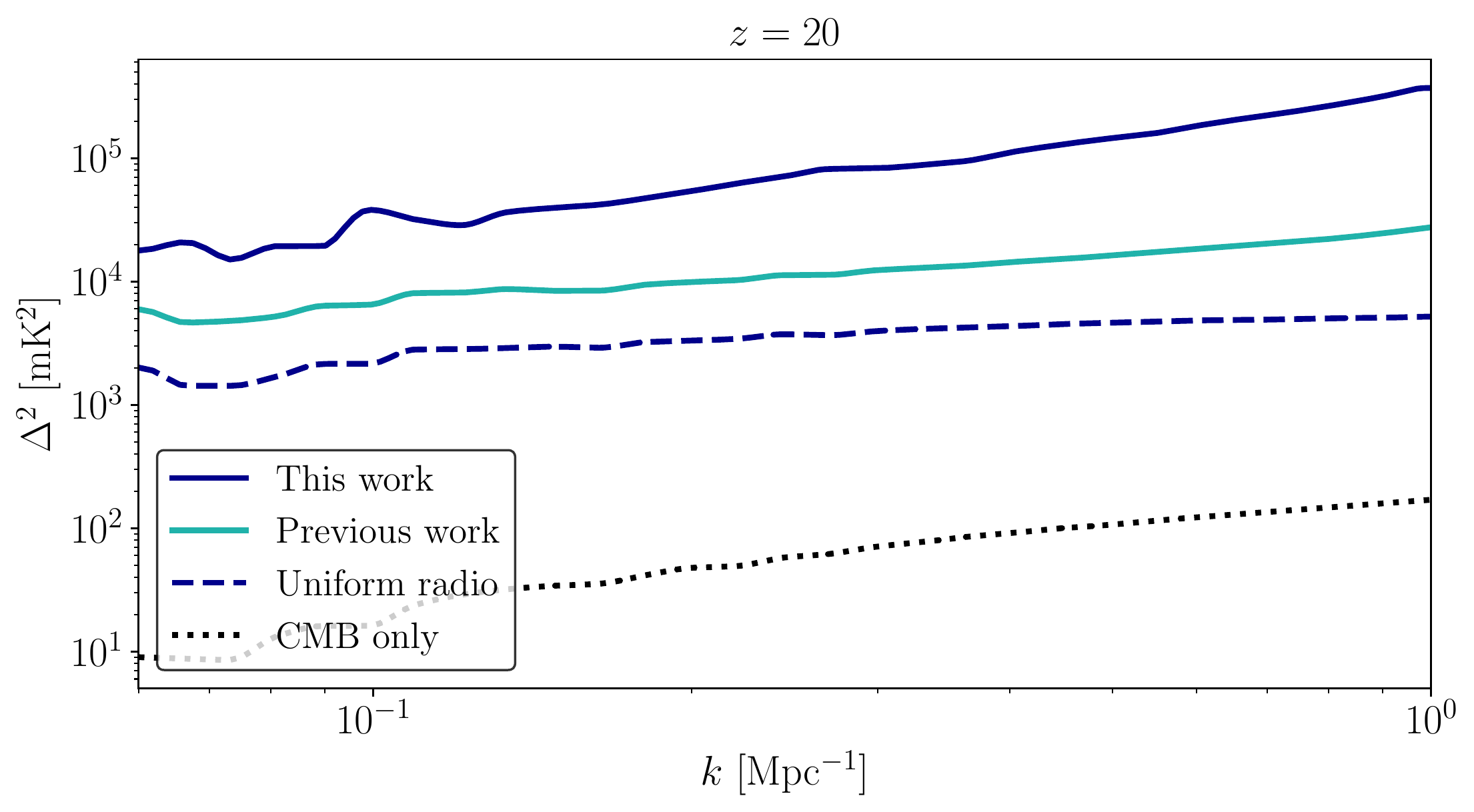}
    \includegraphics[width=0.48\textwidth]{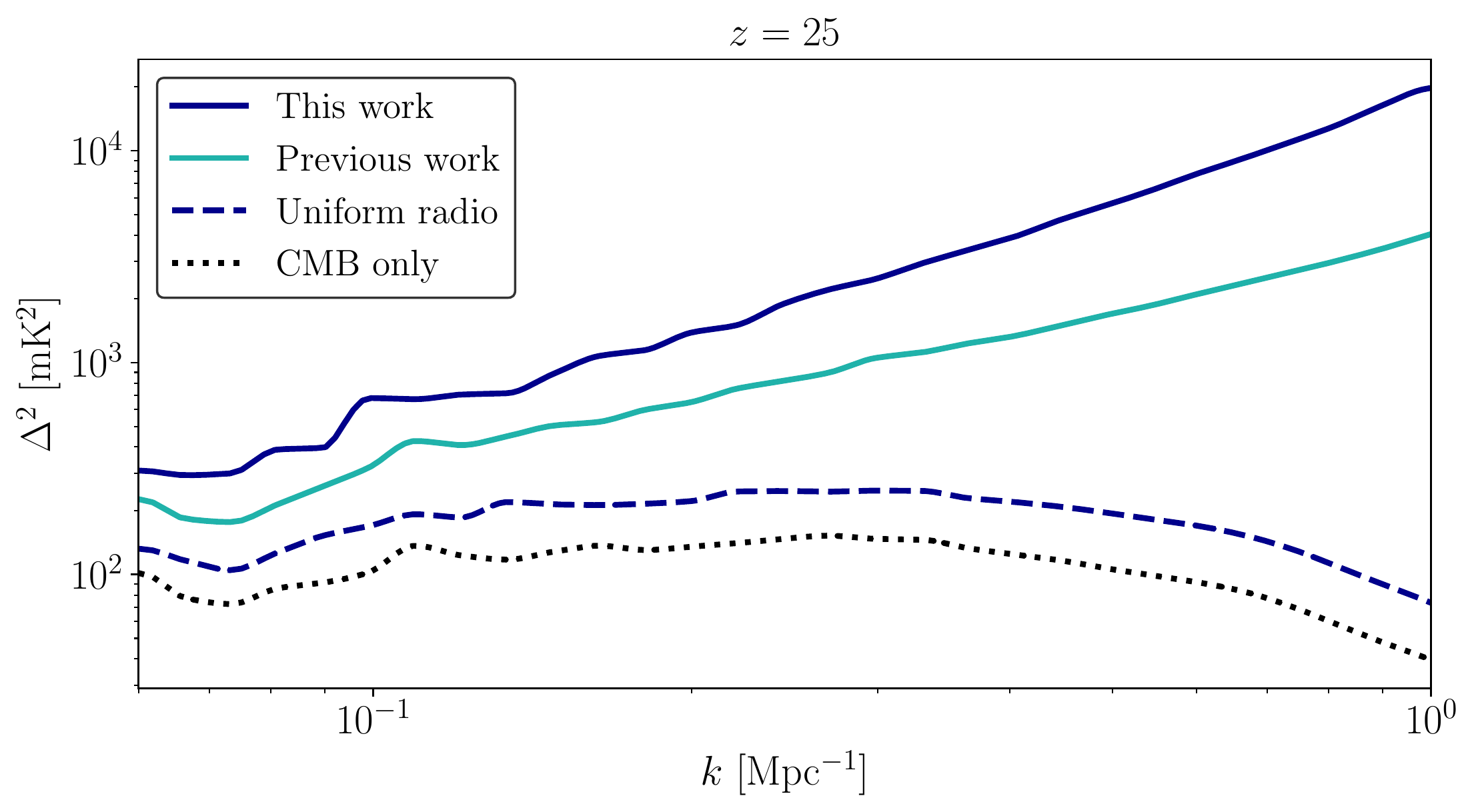}
    \caption{The shape of the 21-cm power spectrum at a given redshift during the epoch of cosmic dawn, comparing previous work with our full calculation that includes the line of sight effect on the excess radio background. The panels show the power spectrum at two different redshifts. Shown here is the case with $f_{\rm{R}} = 3000$ and a hard X-ray SED ($E_{\rm{min}} = 1$ keV) as in Fig. \ref{fig:power_spectrum_z}.}
    \label{fig:power_spectrum_k}
\end{figure*}

Since the (isotropically averaged) 21-cm power spectrum is a function of two variables ($k$ and $z$), in 
Fig.~\ref{fig:power_spectrum_k} we show the other cut, i.e., the function of $k$ at a given redshift. This shows the effect of the LoS radio fluctuations on the shape of the 21-cm power spectrum. Here we use the same astrophysical model parameters as in Fig. \ref{fig:power_spectrum_z}. At high redshifts (right panel of Fig. \ref{fig:power_spectrum_k}),  when the first stars and galaxies begin to form, the Ly-$\alpha$ fluctuations are the dominant source of the 21-cm fluctuations in the standard case. The Ly-$\alpha$ photons typically travel a significant distance~\citep{reis20b}, which washes out small-scale fluctuations, while the strong radio intensity near sources increases small-scale fluctuations, even more when the LoS effect is included. At lower redshift (left panel), the LoS effect has less of an effect on the power spectrum shape, and just gives an overall boost.  At later times, $z\sim 10$, the LoS radio fluctuations have almost no effect on the shape of the 21-cm power spectrum compared to the previous radio fluctuation work \citep{Reis2020} due to the disappearance of the excess radio fluctuations at low redshift. 

\begin{figure*}
    \centering
    \includegraphics[width=0.48\textwidth]{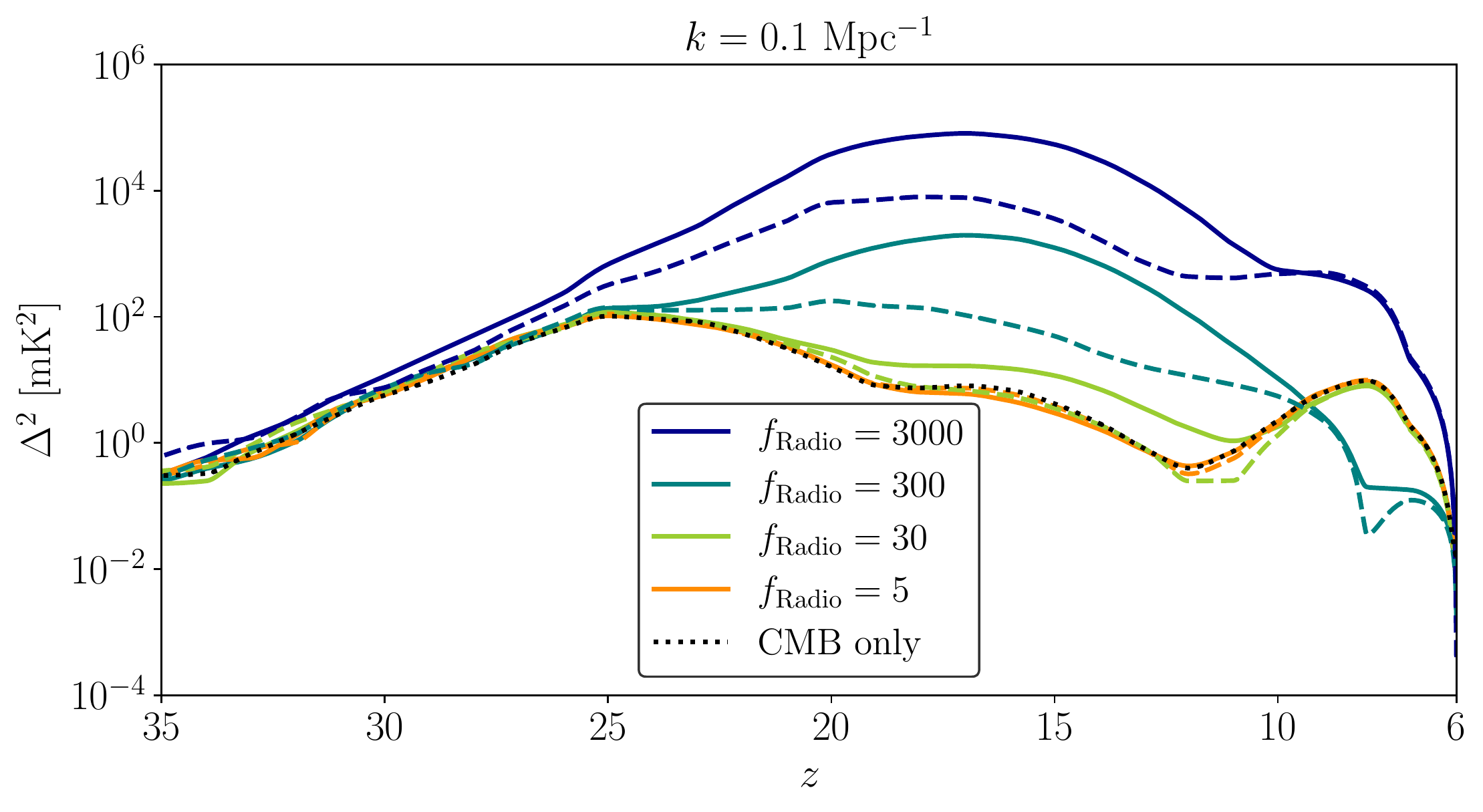}
    \includegraphics[width=0.48\textwidth]{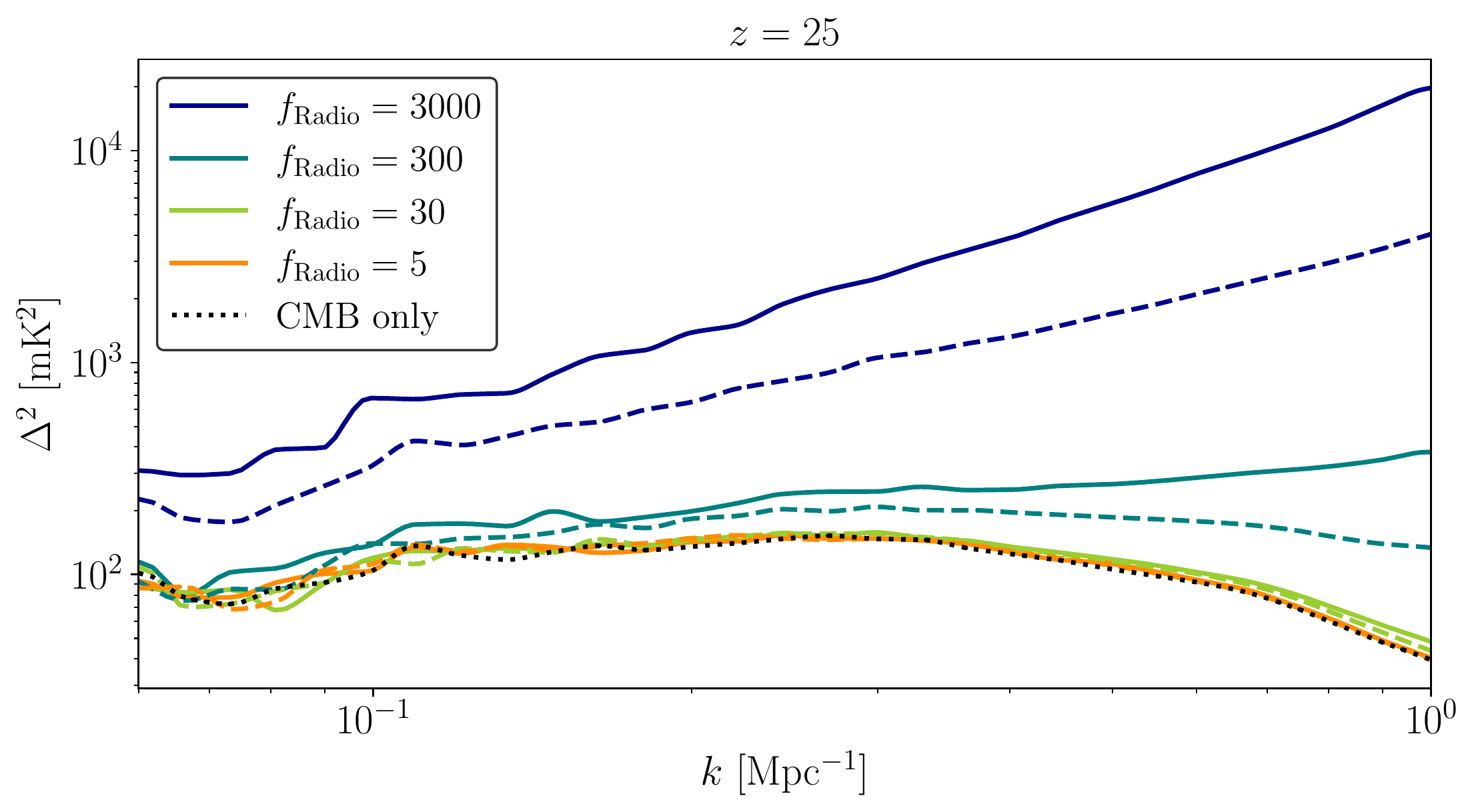}
    \caption{\textbf{Left panel:} The 21-cm power spectrum as a function of redshift at $k = 0.1$ Mpc$^{-1}$ for various values of $f_{\rm{Radio}}$. \textbf{Right panel:} The shape of the 21-cm power spectrum at $z=25$ for various values of $f_{\rm{Radio}}$. We show the full model with the LoS effect on the radio background (solid) compared to the fluctuating radio background (dashed) considered in our previous work \citep{Reis2020}. We also show the case without any excess radio background (i.e., the CMB-only case, black dotted line). Shown here is the model with a hard X-ray SED ($E_{\rm{min}} = 1$ keV).}\label{fig:ps_diff_radio}
\end{figure*}

Up to now in this section we have illustrated the consequences of the LoS effect on the radio background using a particularly strong radio background. Next, we examine the effect of the LoS radio fluctuations on the 21-cm power spectrum while varying the parameter $f_{\rm{Radio}}$ that regulates the strength of the excess radio background. We show the 21-cm power spectrum as a function of redshift at $k = 0.1$ Mpc$^{-1}$ for various values of $f_{\rm{Radio}}$ in Fig. \ref{fig:ps_diff_radio}. The LoS radio fluctuations for moderate values of $f_{\rm{Radio}}$ can have a significant effect ($\sim$ an order of magnitude) after the onset of star formation till the end of the heating transition, as can be seen in the left panel of Fig.\ref{fig:ps_diff_radio}. The right panel of Fig. \ref{fig:ps_diff_radio} shows the shape of the 21-cm power spectrum for various values of $f_{\rm{Radio}}$ at $z = 25$. Among the values in the plot, at high redshift only $f_{\rm{Radio}}$ of at least 300 has a large effect, but later in cosmic dawn even $f_{\rm{Radio}} = 30$ has quite a significant effect; since we set $f_*=0.1$, the latter value corresponds to a value of $f_{\rm{Radio}} \times f_*$ that is lower by two orders of magnitude than the value required to match the EDGES measurement. 

\begin{figure}
    \includegraphics[width=\columnwidth]{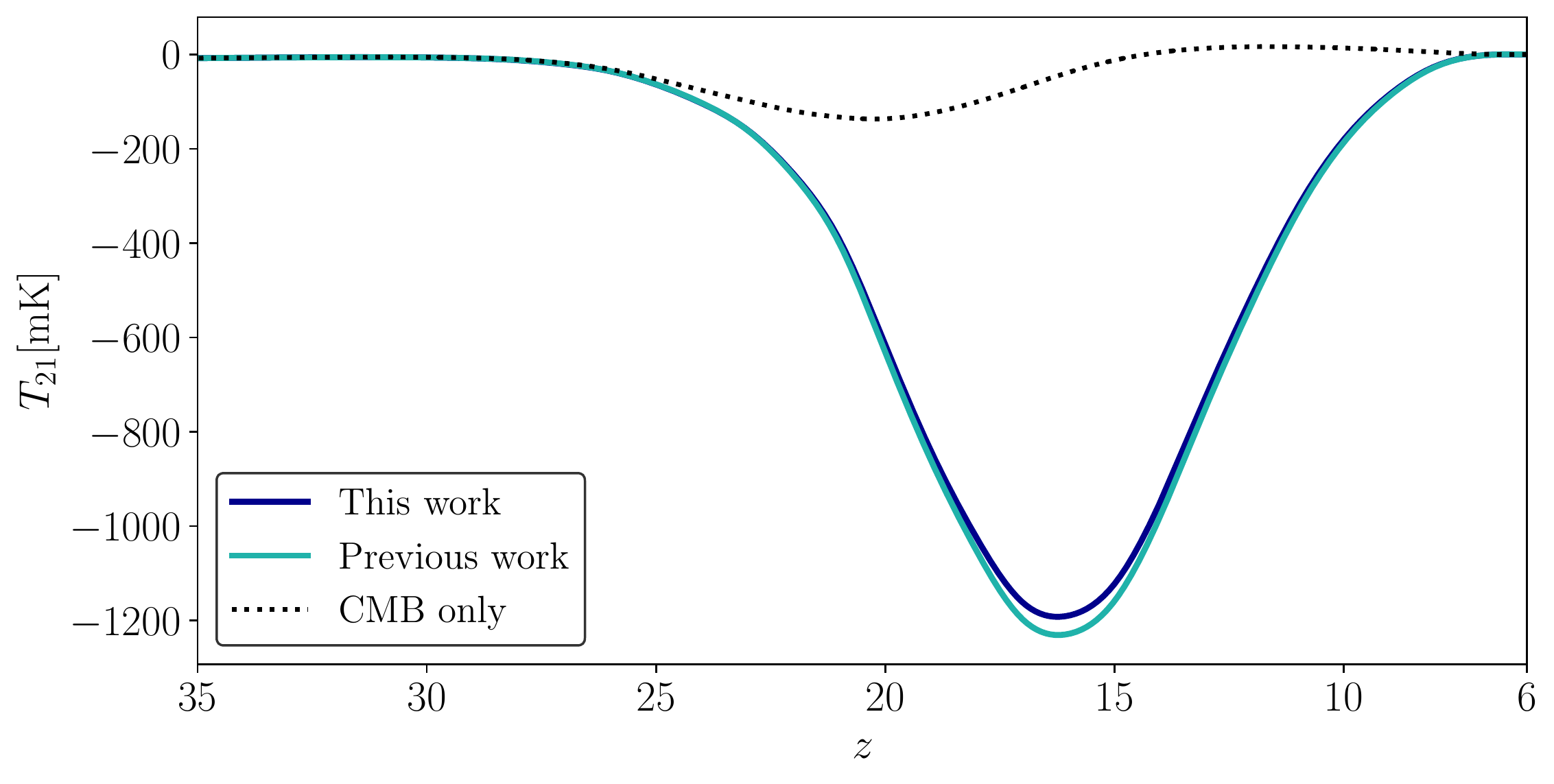}
    \caption{The effect of the radio fluctuations and of the line-of-sight component on the 21-cm global signal, for the astrophysical model of a high radio production efficiency, $f_R = 3000$, and halos with $V_c = 16.5$ km/s, $f_X = 1$, $f_* = 0.1$, and a hard X-ray SED.}
    \label{fig:global_signal}
\end{figure}

The effect of the radio fluctuations on the global 21-cm signal is shown in Fig.~\ref{fig:global_signal}. The LoS effect of the radio fluctuations has only a small effect on the global signal compared to the radio fluctuation model considered in our previous work \citep{Reis2020}. The LoS radio background (solid blue line) results in a slightly shallower minimum (a difference of $\sim 50$ mK). While the mean radio background is unchanged by the LoS effect, the non-linearity of the 21-cm fluctuations causes a slight change in the mean global signal.

\subsection{Quantifying the LoS anisotropy in the 21-cm power spectrum}

In the previous subsection, we illustrated the effect of the LoS radio fluctuations on the 21-cm signal in detail. In this subsection, we analyze and quantify the anisotropy present in the 21-cm power spectrum using the anisotropy ratio of eq.~\ref{eqn:anisotropic_ratio}. This anisotropy includes a) the normal LoS anisotropy due to the radial component of the peculiar velocity gradient \citep{barkana2005}, and b) the anisotropy due to the LoS radio fluctuations. 

\begin{figure}
    \centering
    \includegraphics[width=0.48\textwidth]{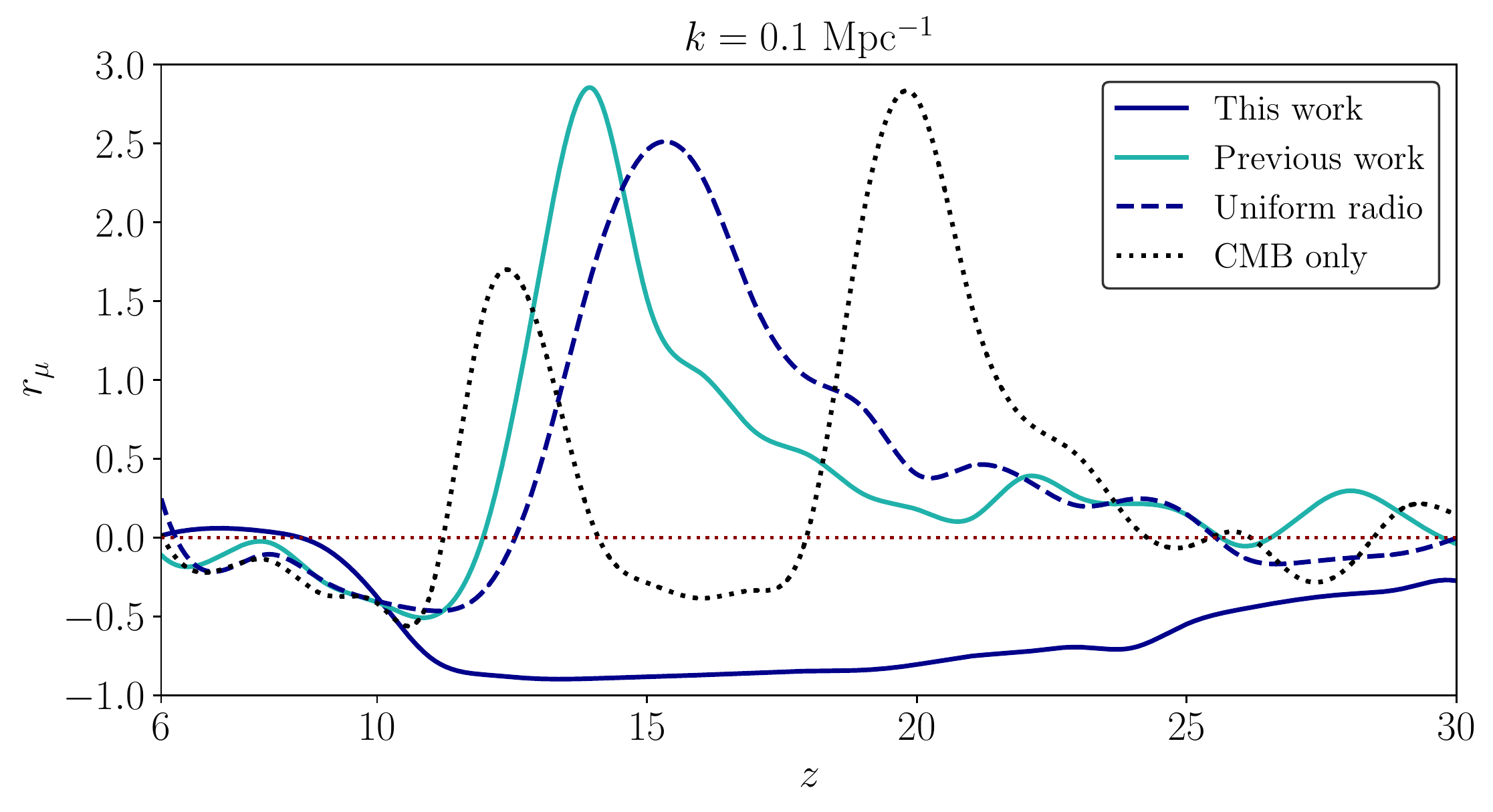}
    \includegraphics[width=0.48\textwidth]{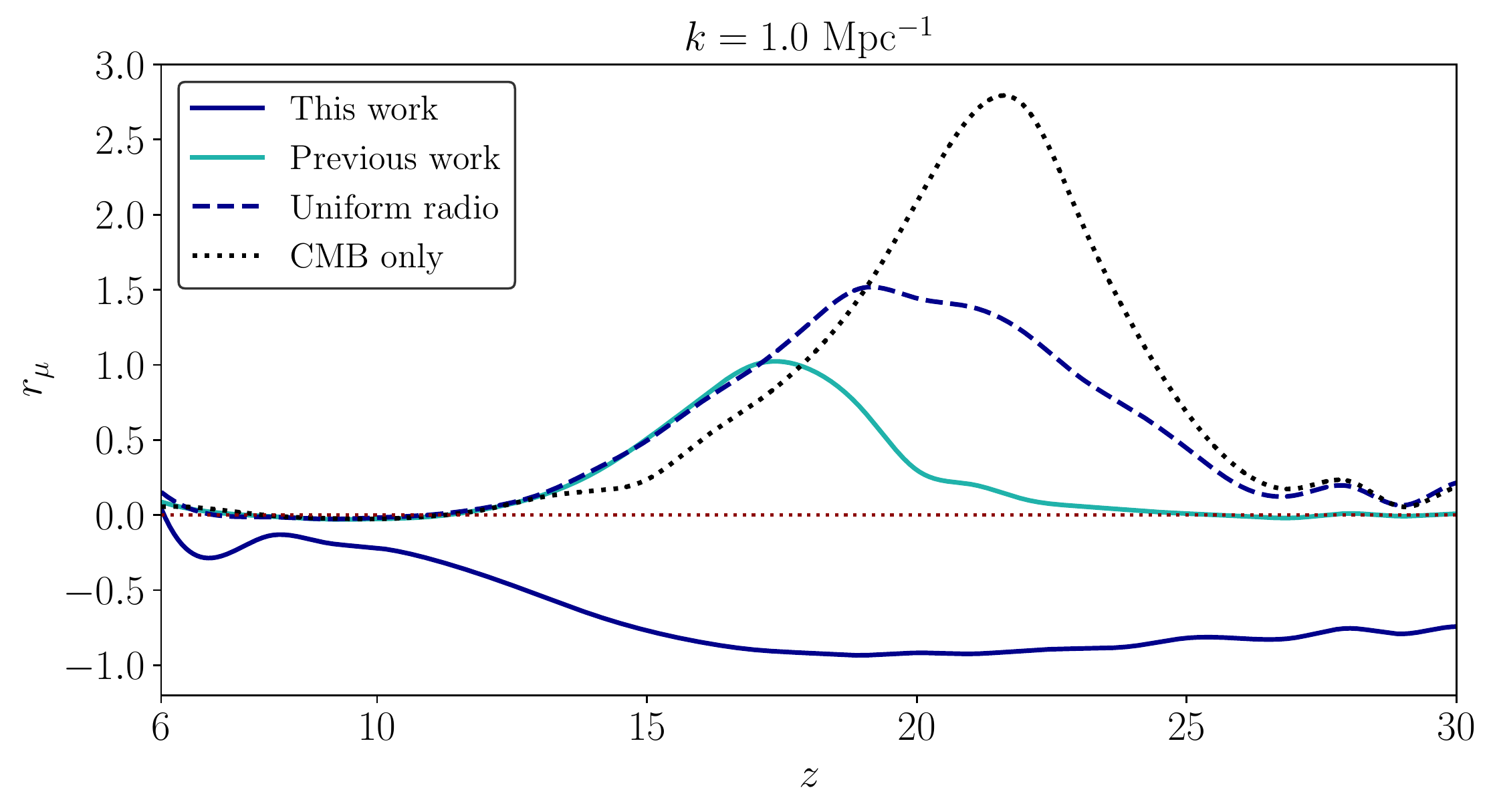}
    \caption{The anisotropy ratio ($r_{\mu}$) of the 21-cm power spectrum as a function of redshift at two wavenumbers: $k = 0.1 \ \rm{Mpc}^{-1}$ (\textbf{top panel}) and $k = 1.0\ \rm{Mpc}^{-1}$ (\textbf{bottom panel}) for various simulation runs: no excess radio or CMB only case (dotted black line), uniform excess radio background (dashed blue line), fluctuating radio background without the LoS effect (solid teal line) and full fluctuating radio background including the LoS effect (solid blue line). The dotted dark red horizontal line indicates $r_{\mu} = 0$.} 
    \label{fig:anisotropy_z}
\end{figure}

We show in Fig. \ref{fig:anisotropy_z} the angular dependence of the 21-cm power spectrum as a function of redshift for two wavenumbers: $k = 0.1 \ \rm{Mpc}^{-1}$ (top panel) and $k = 1.0\ \rm{Mpc}^{-1}$ (bottom panel). The dotted black line represents the case with no excess radio background (CMB only case), in wh ich the anisotropy showing is only due to the LoS peculiar velocity gradient. In the top panel of Fig. \ref{fig:anisotropy_z}, we see two clear peaks (in the dotted black line) at $z \sim 20-21$ (when the Ly-$\alpha$ fluctuations dominate the 21-cm signal) and $z \sim 12-13$ (during the heating transition). The colored lines show the cases when we include the excess radio background from previous work (solid teal line), the radio background with the LoS effect from this work (solid blue line), and the comparison case of a uniform excess radio background (dashed blue line). The case of a uniform radio background shows a single peak of $r_{\mu}$ at $z \sim 15-16$, while the case of the fluctuating radio background from the previous work also shows a single peak in $r_{\mu}$, at a slightly lower redshift ($z \sim 14$). In the bottom panel of Fig.~\ref{fig:anisotropy_z}, these three cases again show a single significant positive peak of $r_{\mu}$, at a redshift between 17 and 22. 

\begin{figure}
\includegraphics[width=0.48\textwidth]{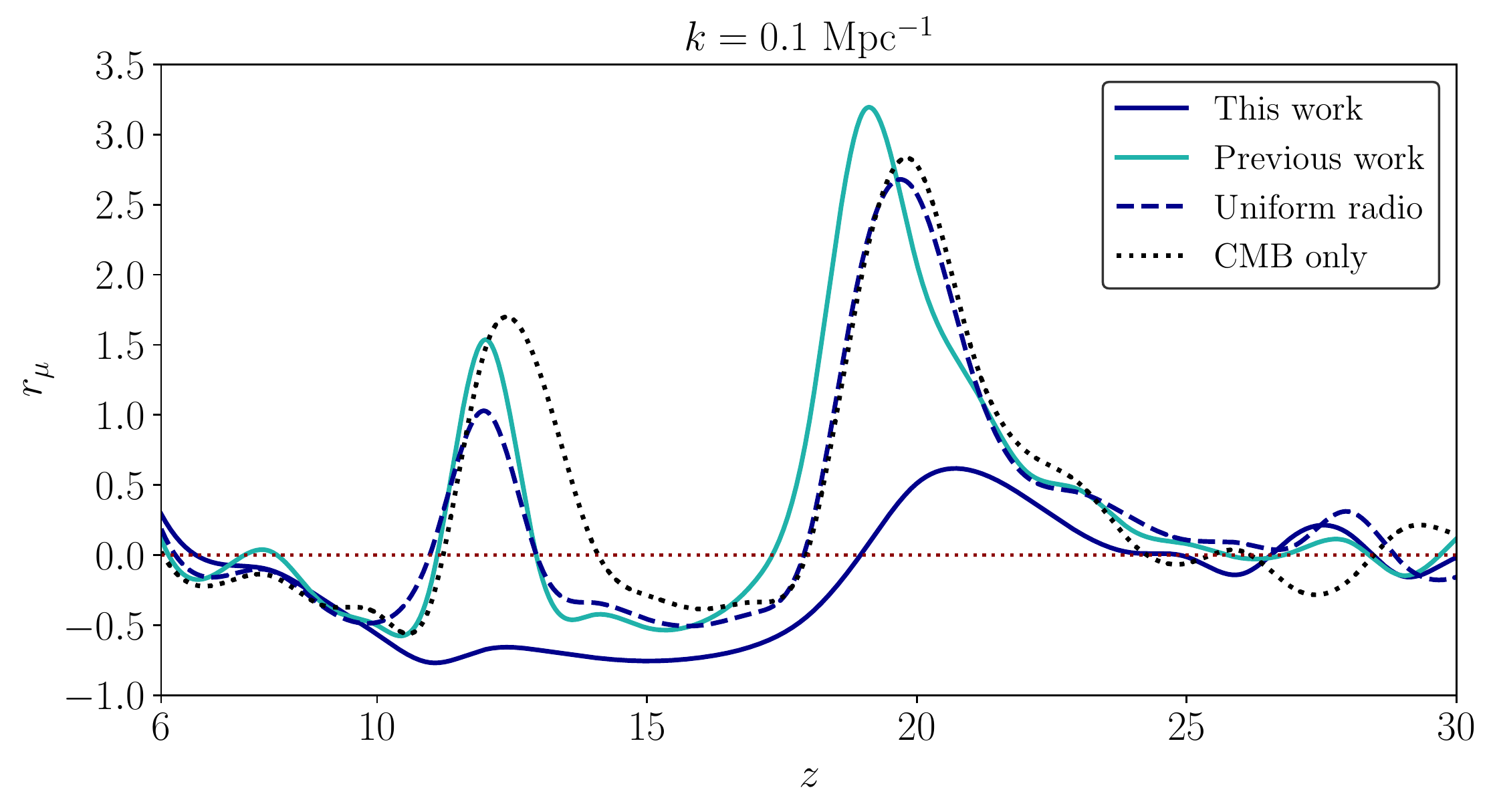}
\includegraphics[width=0.48\textwidth]{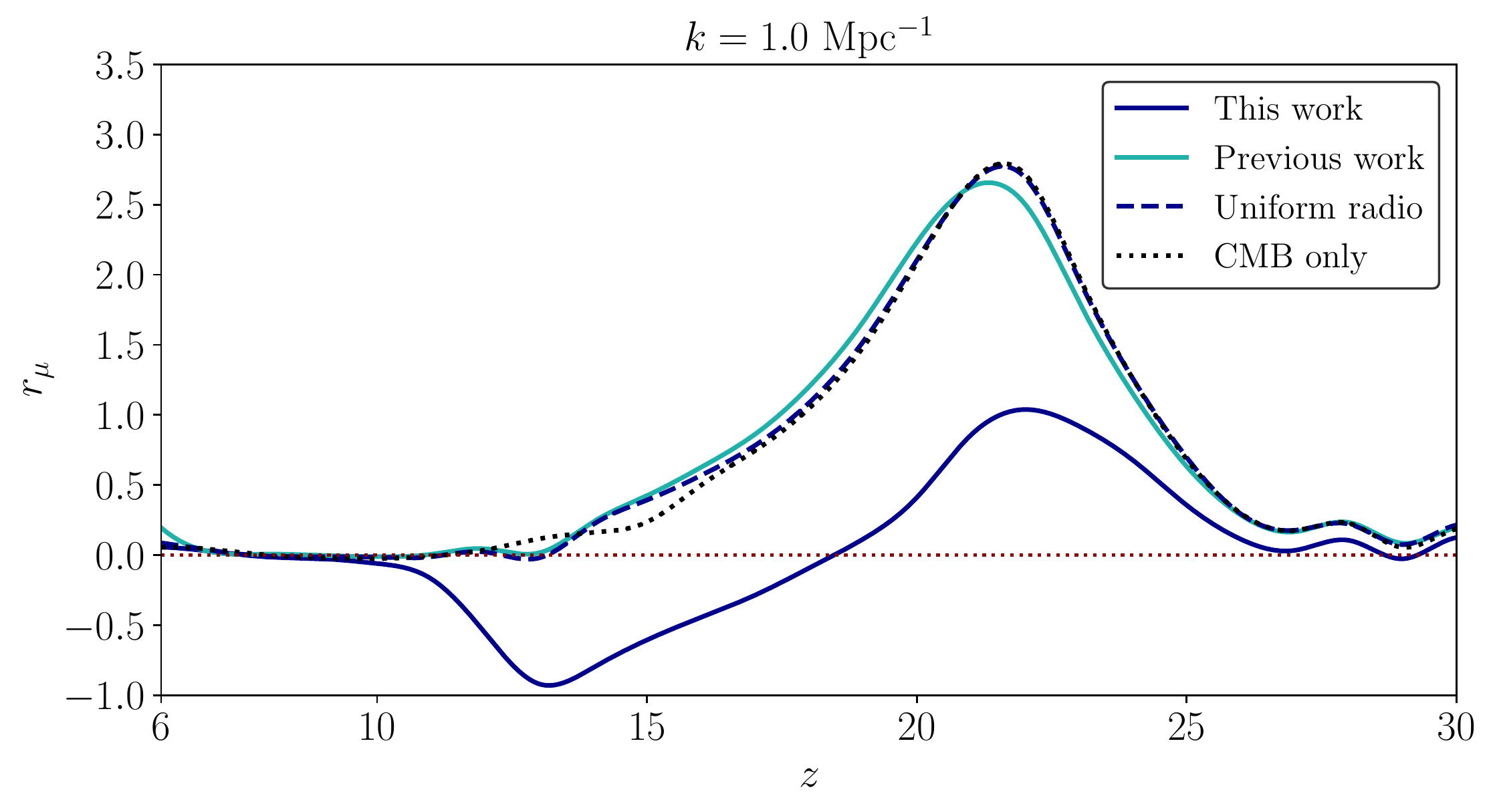}
\caption{The anisotropy ratio ($r_{\mu}$) of the 21-cm power spectrum. Same as Fig.~\ref{fig:anisotropy_z} but for $f_{\rm{Radio}} = 30$.}
 \label{fig:anisotropy_fR}
\end{figure}

In contrast, the LoS effect on the radio fluctuations (solid blue lines in the two panels of Fig.~\ref{fig:anisotropy_z}) completely changes the anisotropy. It makes the anisotropy ratio $r_{\mu}$ negative (even approaching its lowest possible value of $-1$) throughout cosmic dawn, until late in the epoch of reionization. The reason is that each radio source lights up a pencil beam between us and the source (see the right-most panel of Fig.~\ref{fig:21cm_slice_hard}). This smooths out the fluctuations along the line of sight, making them small compared to those that are perpendicular to the LoS. Fig.~\ref{fig:anisotropy_fR} shows the much smaller value of $f_{\rm{Radio}} = 30$, corresponding to moderate enhancement of the radio background. In this case, the anisotropy ratio would be only slightly changed by the radio background without the LoS effect, but it is still substantially lowered by the LoS effect; it goes down to negative values at $z$ below 18, and the peak at $z \sim 12-13$ is erased for $k = 0.1 \ \rm{Mpc}^{-1}$, while a negative peak is created for $k = 1 \ \rm{Mpc}^{-1}$.

\begin{figure*}
\centering
\includegraphics[width=0.96\textwidth]{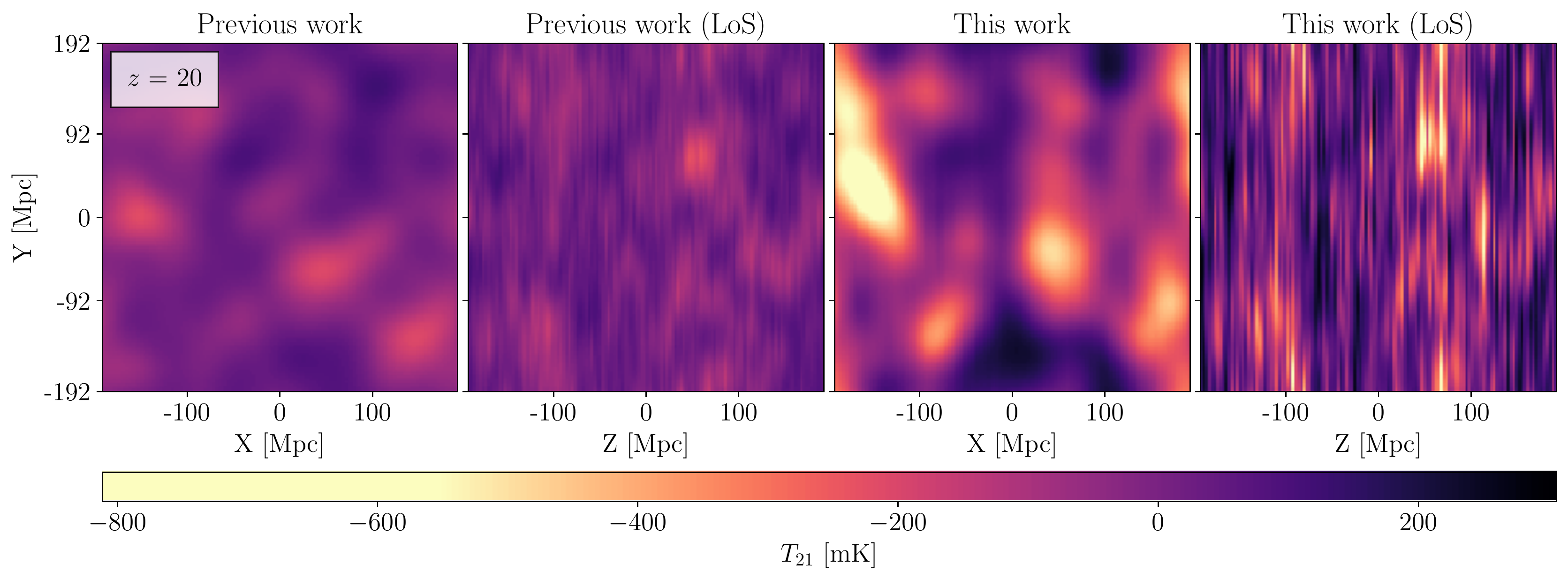}
\caption{Mock SKA images of the cosmic dawn 21-cm signal from $z=20$. The panels show the same simulated slices as in Fig.~\ref{fig:21cm_slice_hard}, but with several observational effects, simulated as expected for the SKA: angular resolution, thermal noise, and foreground avoidance.}
\label{fig:21cm_SKA}
\end{figure*}

These results show that the anisotropy yields a potentially clear observational signature of the LoS effect of bright early radio sources. However, we must add a note of caution. These are theoretical, simulated results, while in practice observations face additional obstacles. Specifically, in any 21-cm measurement, the strong radio synchrotron foreground (including the emission coming from the Milky Way) has a rather smooth spectrum, so foreground removal will likely eliminate much of the pencil beam signature seen in Fig.~\ref{fig:21cm_slice_hard}. To illustrate this, we show in Fig.~\ref{fig:21cm_SKA} a realistic observational version (corresponding to expectations for the SKA) of Fig.~\ref{fig:21cm_slice_hard}; following \citet{reis20b} (see that reference for details), we include angular smoothing corresponding to the angular resolution, thermal noise, and foreground avoidance corresponding to the removal of a foreground-dominated wedge in $\bf{k}$ space. Comparing the $XY$ slices, we see that the LoS effect on the excess radio background produces a strong enhancement of the bright regions in the SKA images (corresponding to galaxy concentrations). Comparing the $ZY$ slices, we again see the enhancement in the rightmost panel with the LoS effect, but while the angular smoothing is seen in the $Y$ direction, the pencil beam features in the $Z$ direction are not apparent. We leave for future work a quantitative assessment of methods for detecting the anisotropy that account for the need for foreground avoidance.

\section{Summary}\label{sec:summary}

In this paper, we showed that individual radio sources acting as background 21-cm sources along the line of sight significantly change the expected 21-cm signal, if early galaxies were particularly bright in low-frequency radio emission. In particular, we found that the LoS effect on the radio fluctuations boosts the 21-cm power spectrum throughout cosmic dawn, by an additional order of magnitude beyond just the enhanced radio background and its angle-averaged fluctuations. The radio fluctuations, enhanced by the LoS effect, wash out the Ly-$\alpha$ and heating peaks and produce a single broad peak at cosmic dawn, when the power spectrum is considered as a function of redshift; this is true at $k=0.1$ Mpc$^{-1}$ as well as 1 Mpc$^{-1}$, and whether the SED of the X-ray heating sources is hard or soft. The LoS radio fluctuations have almost no effect at the end of the epoch of reionization due to the fact that the radio sources become numerous and the radio fluctuations disappear at lower redshifts. The radio fluctuations also change the shape of the 21-cm power spectrum, increasing small-scale fluctuations at the higher redshifts (particular around $z=25$), even more when the LoS effect is included.

We also explored the LoS effect on the radio fluctuations for moderate values of radio enhancement, that are well below the values that are required to explain the EDGES feature. We found that even models with $f_{\rm{Radio}} = 30$ can significantly enhance the 21-cm power spectrum (by a half to one order of magnitude) during most of cosmic dawn. When modelling and interpreting the 21-cm signal, it is therefore important to consider a possible enhanced radio background, and to include the LoS effect. We note that the LoS radio fluctuations also slightly affect the global signal due to the non-linearity of the 21-cm fluctuations.

Finally, we quantified the anisotropy in the 21-cm power spectrum and showed that the LoS effect on the radio background introduces a new anisotropy in the 21-cm power spectrum. The LoS effect specifically produces a negative anisotropy ratio almost over a wide redshift range, at least in the case of a very high (EDGES motivated) radio production efficiency ($f_{R} = 3000$).  Even a moderate radio efficiency ($f_{\rm{Radio}} = 30$) still produces a unique signature in the anisotropy ratio, with the LoS effect producing a negative anisotropy ratio
for part of cosmic dawn. However, we caution that further investigation is needed to see whether the anisotropy ratio can be measured in the case of a realistic power spectrum, such as that expected from SKA observations. 
\section*{Acknowledgements}

SS and RB acknowledge the support of the Israel Science Foundation (grant No. 2359/20). RB also thanks the Ambrose Monell Foundation and the Institute for Advanced Study as well as the Vera Rubin Presidential Chair in Astronomy at UCSC and the Packard Foundation. AF was supported by the Royal Society University Research Fellowship. This research made use of: \texttt{Numpy} \citep{harris2020array}, \texttt{Scipy} \citep{2020SciPy-NMeth}, \texttt{matplotlib} \citep{Hunter:2007}
and the NASA Astrophysics Data System Bibliographic Services.
\section*{Data Availability}

The data underlying this article will be shared on reasonable request to the corresponding author.



\bibliographystyle{mnras}
\bibliography{example} 

\begin{thebibliography}{}
\makeatletter
\relax
\def\mn@urlcharsother{\let\do\@makeother \do\$\do\&\do\#\do\^\do\_\do\%\do\~}
\def\mn@doi{\begingroup\mn@urlcharsother \@ifnextchar [ {\mn@doi@}
  {\mn@doi@[]}}
\def\mn@doi@[#1]#2{\def\@tempa{#1}\ifx\@tempa\@empty \href
  {http://dx.doi.org/#2} {doi:#2}\else \href {http://dx.doi.org/#2} {#1}\fi
  \endgroup}
\def\mn@eprint#1#2{\mn@eprint@#1:#2::\@nil}
\def\mn@eprint@arXiv#1{\href {http://arxiv.org/abs/#1} {{\tt arXiv:#1}}}
\def\mn@eprint@dblp#1{\href {http://dblp.uni-trier.de/rec/bibtex/#1.xml}
  {dblp:#1}}
\def\mn@eprint@#1:#2:#3:#4\@nil{\def\@tempa {#1}\def\@tempb {#2}\def\@tempc
  {#3}\ifx \@tempc \@empty \let \@tempc \@tempb \let \@tempb \@tempa \fi \ifx
  \@tempb \@empty \def\@tempb {arXiv}\fi \@ifundefined
  {mn@eprint@\@tempb}{\@tempb:\@tempc}{\expandafter \expandafter \csname
  mn@eprint@\@tempb\endcsname \expandafter{\@tempc}}}

\bibitem[\protect\citeauthoryear{{Alcock} \& {Paczy\'{n}ski}}{{Alcock} \&
  {Paczy\'{n}ski}}{1979}]{alcock_paczynski}
{Alcock} C.,  {Paczy\'{n}ski} B.,  1979, \mn@doi [\nat] {10.1038/281358a0},
  \href {https://ui.adsabs.harvard.edu/abs/1979Natur.281..358A} {281, 358}

\bibitem[\protect\citeauthoryear{{Ali}, {Bharadwaj}  \& {Pandey}}{{Ali}
  et~al.}{2005}]{APindian}
{Ali} S.~S.,  {Bharadwaj} S.,   {Pandey} B.,  2005, \mn@doi [\mnras]
  {10.1111/j.1365-2966.2005.09444.x}, \href
  {https://ui.adsabs.harvard.edu/abs/2005MNRAS.363..251A} {363, 251}

\bibitem[\protect\citeauthoryear{{Allison} \& {Dalgarno}}{{Allison} \&
  {Dalgarno}}{1969}]{allison69}
{Allison} A.~C.,  {Dalgarno} A.,  1969, \mn@doi [\apj] {10.1086/150204}, \href
  {https://ui.adsabs.harvard.edu/abs/1969ApJ...158..423A} {158, 423}

\bibitem[\protect\citeauthoryear{{Barkana}}{{Barkana}}{2006}]{MeAP}
{Barkana} R.,  2006, \mn@doi [\mnras] {10.1111/j.1365-2966.2006.10882.x}, \href
  {https://ui.adsabs.harvard.edu/abs/2006MNRAS.372..259B} {372, 259}

\bibitem[\protect\citeauthoryear{{Barkana}}{{Barkana}}{2018}]{barkana18}
{Barkana} R.,  2018, \mn@doi [\nat] {10.1038/nature25791}, \href
  {https://ui.adsabs.harvard.edu/abs/2018Natur.555...71B} {555, 71}

\bibitem[\protect\citeauthoryear{{Barkana} \& {Loeb}}{{Barkana} \&
  {Loeb}}{2004}]{barkana04}
{Barkana} R.,  {Loeb} A.,  2004, \mn@doi [\apj] {10.1086/421079}, \href
  {https://ui.adsabs.harvard.edu/abs/2004ApJ...609..474B} {609, 474}

\bibitem[\protect\citeauthoryear{{Barkana} \& {Loeb}}{{Barkana} \&
  {Loeb}}{2005}]{barkana2005}
{Barkana} R.,  {Loeb} A.,  2005, \mn@doi [\apjl] {10.1086/430599}, \href
  {https://ui.adsabs.harvard.edu/abs/2005ApJ...624L..65B} {624, L65}

\bibitem[\protect\citeauthoryear{{Barkana} \& {Loeb}}{{Barkana} \&
  {Loeb}}{2006}]{barkana_loeb2006}
{Barkana} R.,  {Loeb} A.,  2006, \mn@doi [\mnras]
  {10.1111/j.1745-3933.2006.00222.x}, \href
  {https://ui.adsabs.harvard.edu/abs/2006MNRAS.372L..43B} {372, L43}

\bibitem[\protect\citeauthoryear{{Barkana}, {Outmezguine}, {Redigol}  \&
  {Volansky}}{{Barkana} et~al.}{2018}]{barkana18a}
{Barkana} R.,  {Outmezguine} N.~J.,  {Redigol} D.,   {Volansky} T.,  2018,
  \mn@doi [\prd] {10.1103/PhysRevD.98.103005}, \href
  {https://ui.adsabs.harvard.edu/abs/2018PhRvD..98j3005B} {98, 103005}

\bibitem[\protect\citeauthoryear{{Barkana}, {Fialkov}, {Liu}  \&
  {Outmezguine}}{{Barkana} et~al.}{2022}]{Barkana2022}
{Barkana} R.,  {Fialkov} A.,  {Liu} H.,   {Outmezguine} N.~J.,  2022, arXiv
  e-prints, \href {https://ui.adsabs.harvard.edu/abs/2022arXiv221208082B} {p.
  arXiv:2212.08082}

\bibitem[\protect\citeauthoryear{Berlin, Hooper, Krnjaic  \& McDermott}{Berlin
  et~al.}{2018}]{Berlin}
Berlin A.,  Hooper D.,  Krnjaic G.,   McDermott S.~D.,  2018, \mn@doi [Phys.
  Rev. Lett.] {10.1103/PhysRevLett.121.011102}, 121, 011102

\bibitem[\protect\citeauthoryear{{Bharadwaj} \& {Ali}}{{Bharadwaj} \&
  {Ali}}{2004}]{SB_SSA_2004}
{Bharadwaj} S.,  {Ali} S.~S.,  2004, \mn@doi [\mnras]
  {10.1111/j.1365-2966.2004.07907.x}, \href
  {https://ui.adsabs.harvard.edu/abs/2004MNRAS.352..142B} {352, 142}

\bibitem[\protect\citeauthoryear{{Biermann}, {Nath}, {Caramete}, {Harms},
  {Stanev}  \& {Becker Tjus}}{{Biermann} et~al.}{2014}]{biermann14}
{Biermann} P.~L.,  {Nath} B.~B.,  {Caramete} L.~I.,  {Harms} B.~C.,  {Stanev}
  T.,   {Becker Tjus} J.,  2014, \mn@doi [\mnras] {10.1093/mnras/stu541}, \href
  {https://ui.adsabs.harvard.edu/abs/2014MNRAS.441.1147B} {441, 1147}

\bibitem[\protect\citeauthoryear{{Bolgar}, {Eames}, {Hottier}  \&
  {Semelin}}{{Bolgar} et~al.}{2018}]{bolgar18}
{Bolgar} F.,  {Eames} E.,  {Hottier} C.,   {Semelin} B.,  2018, \mn@doi
  [\mnras] {10.1093/mnras/sty1293}, \href
  {https://ui.adsabs.harvard.edu/abs/2018MNRAS.478.5564B} {478, 5564}

\bibitem[\protect\citeauthoryear{{Bowman}, {Rogers}, {Monsalve}, {Mozdzen}  \&
  {Mahesh}}{{Bowman} et~al.}{2018}]{bowman18}
{Bowman} J.~D.,  {Rogers} A. E.~E.,  {Monsalve} R.~A.,  {Mozdzen} T.~J.,
  {Mahesh} N.,  2018, \mn@doi [\nat] {10.1038/nature25792}, \href
  {https://ui.adsabs.harvard.edu/abs/2018Natur.555...67B} {555, 67}

\bibitem[\protect\citeauthoryear{{Brandenberger}, {Cyr}  \&
  {Shi}}{{Brandenberger} et~al.}{2019}]{Brandenberger:2019}
{Brandenberger} R.,  {Cyr} B.,   {Shi} R.,  2019, \mn@doi [\jcap]
  {10.1088/1475-7516/2019/09/009}, \href
  {https://ui.adsabs.harvard.edu/abs/2019JCAP...09..009B} {2019, 009}

\bibitem[\protect\citeauthoryear{{Cohen}, {Fialkov}  \& {Barkana}}{{Cohen}
  et~al.}{2016}]{cohen16}
{Cohen} A.,  {Fialkov} A.,   {Barkana} R.,  2016, \mn@doi [\mnras]
  {10.1093/mnrasl/slw047}, \href
  {https://ui.adsabs.harvard.edu/abs/2016MNRAS.459L..90C} {459, L90}

\bibitem[\protect\citeauthoryear{{Cohen}, {Fialkov}, {Barkana}  \&
  {Lotem}}{{Cohen} et~al.}{2017}]{cohen17}
{Cohen} A.,  {Fialkov} A.,  {Barkana} R.,   {Lotem} M.,  2017, \mn@doi [\mnras]
  {10.1093/mnras/stx2065}, \href
  {https://ui.adsabs.harvard.edu/abs/2017MNRAS.472.1915C} {472, 1915}

\bibitem[\protect\citeauthoryear{{Condon}}{{Condon}}{1992}]{condon92}
{Condon} J.~J.,  1992, \mn@doi [\araa] {10.1146/annurev.aa.30.090192.003043},
  \href {https://ui.adsabs.harvard.edu/abs/1992ARA&A..30..575C} {30, 575}

\bibitem[\protect\citeauthoryear{{Condon}, {Cotton}  \& {Broderick}}{{Condon}
  et~al.}{2002}]{condon02}
{Condon} J.~J.,  {Cotton} W.~D.,   {Broderick} J.~J.,  2002, \mn@doi [\aj]
  {10.1086/341650}, \href
  {https://ui.adsabs.harvard.edu/abs/2002AJ....124..675C} {124, 675}

\bibitem[\protect\citeauthoryear{{Datta}, {Mellema}, {Mao}, {Iliev}, {Shapiro}
  \& {Ahn}}{{Datta} et~al.}{2012}]{Datta12}
{Datta} K.~K.,  {Mellema} G.,  {Mao} Y.,  {Iliev} I.~T.,  {Shapiro} P.~R.,
  {Ahn} K.,  2012, \mn@doi [\mnras] {10.1111/j.1365-2966.2012.21293.x}, \href
  {https://ui.adsabs.harvard.edu/abs/2012MNRAS.424.1877D} {424, 1877}

\bibitem[\protect\citeauthoryear{{Dowell} \& {Taylor}}{{Dowell} \&
  {Taylor}}{2018}]{dowell18}
{Dowell} J.,  {Taylor} G.~B.,  2018, \mn@doi [\apjl]
  {10.3847/2041-8213/aabf86}, \href
  {https://ui.adsabs.harvard.edu/abs/2018ApJ...858L...9D} {858, L9}

\bibitem[\protect\citeauthoryear{{Ewall-Wice}, {Chang}, {Lazio}, {Dor{\'e}},
  {Seiffert}  \& {Monsalve}}{{Ewall-Wice} et~al.}{2018}]{ewall18}
{Ewall-Wice} A.,  {Chang} T.~C.,  {Lazio} J.,  {Dor{\'e}} O.,  {Seiffert} M.,
  {Monsalve} R.~A.,  2018, \mn@doi [\apj] {10.3847/1538-4357/aae51d}, \href
  {https://ui.adsabs.harvard.edu/abs/2018ApJ...868...63E} {868, 63}

\bibitem[\protect\citeauthoryear{{Ewall-Wice}, {Chang}  \&
  {Lazio}}{{Ewall-Wice} et~al.}{2020}]{ewall20}
{Ewall-Wice} A.,  {Chang} T.-C.,   {Lazio} T. J.~W.,  2020, \mn@doi [\mnras]
  {10.1093/mnras/stz3501}, \href
  {https://ui.adsabs.harvard.edu/abs/2020MNRAS.492.6086E} {492, 6086}

\bibitem[\protect\citeauthoryear{{Feng} \& {Holder}}{{Feng} \&
  {Holder}}{2018}]{feng18}
{Feng} C.,  {Holder} G.,  2018, \mn@doi [\apjl] {10.3847/2041-8213/aac0fe},
  \href {https://ui.adsabs.harvard.edu/abs/2018ApJ...858L..17F} {858, L17}

\bibitem[\protect\citeauthoryear{{Fialkov} \& {Barkana}}{{Fialkov} \&
  {Barkana}}{2014}]{fialkov14}
{Fialkov} A.,  {Barkana} R.,  2014, \mn@doi [\mnras] {10.1093/mnras/stu1744},
  \href {https://ui.adsabs.harvard.edu/abs/2014MNRAS.445..213F} {445, 213}

\bibitem[\protect\citeauthoryear{{Fialkov} \& {Barkana}}{{Fialkov} \&
  {Barkana}}{2019}]{fialkov19}
{Fialkov} A.,  {Barkana} R.,  2019, \mn@doi [\mnras] {10.1093/mnras/stz873},
  \href {https://ui.adsabs.harvard.edu/abs/2019MNRAS.486.1763F} {486, 1763}

\bibitem[\protect\citeauthoryear{{Fialkov}, {Barkana}, {Tseliakhovich}  \&
  {Hirata}}{{Fialkov} et~al.}{2012}]{fialkov12}
{Fialkov} A.,  {Barkana} R.,  {Tseliakhovich} D.,   {Hirata} C.~M.,  2012,
  \mn@doi [\mnras] {10.1111/j.1365-2966.2012.21318.x}, \href
  {https://ui.adsabs.harvard.edu/abs/2012MNRAS.424.1335F} {424, 1335}

\bibitem[\protect\citeauthoryear{{Fialkov}, {Barkana}, {Visbal},
  {Tseliakhovich}  \& {Hirata}}{{Fialkov} et~al.}{2013}]{fialkov2013}
{Fialkov} A.,  {Barkana} R.,  {Visbal} E.,  {Tseliakhovich} D.,   {Hirata}
  C.~M.,  2013, \mn@doi [\mnras] {10.1093/mnras/stt650}, \href
  {https://ui.adsabs.harvard.edu/abs/2013MNRAS.432.2909F} {432, 2909}

\bibitem[\protect\citeauthoryear{{Fialkov}, {Barkana}  \& {Visbal}}{{Fialkov}
  et~al.}{2014}]{fialkov14a}
{Fialkov} A.,  {Barkana} R.,   {Visbal} E.,  2014, \mn@doi [\nat]
  {10.1038/nature12999}, \href
  {https://ui.adsabs.harvard.edu/abs/2014Natur.506..197F} {506, 197}

\bibitem[\protect\citeauthoryear{{Fialkov}, {Barkana}  \& {Cohen}}{{Fialkov}
  et~al.}{2015}]{fialkov2015}
{Fialkov} A.,  {Barkana} R.,   {Cohen} A.,  2015, \mn@doi [\prl]
  {10.1103/PhysRevLett.114.101303}, \href
  {https://ui.adsabs.harvard.edu/abs/2015PhRvL.114j1303F} {114, 101303}

\bibitem[\protect\citeauthoryear{{Fixsen} et~al.,}{{Fixsen}
  et~al.}{2011}]{fixsen11}
{Fixsen} D.~J.,  et~al., 2011, \mn@doi [\apj] {10.1088/0004-637X/734/1/5},
  \href {https://ui.adsabs.harvard.edu/abs/2011ApJ...734....5F} {734, 5}

\bibitem[\protect\citeauthoryear{{Fragos} et~al.,}{{Fragos}
  et~al.}{2013}]{fragos13}
{Fragos} T.,  et~al., 2013, \mn@doi [\apj] {10.1088/0004-637X/764/1/41}, \href
  {https://ui.adsabs.harvard.edu/abs/2013ApJ...764...41F} {764, 41}

\bibitem[\protect\citeauthoryear{{Fraser} et~al.,}{{Fraser}
  et~al.}{2018}]{Fraser:2018}
{Fraser} S.,  et~al., 2018, \mn@doi [Physics Letters B]
  {10.1016/j.physletb.2018.08.035}, \href
  {https://ui.adsabs.harvard.edu/abs/2018PhLB..785..159F} {785, 159}

\bibitem[\protect\citeauthoryear{{Furlanetto}, {Zaldarriaga}  \&
  {Hernquist}}{{Furlanetto} et~al.}{2004}]{furlanetto04}
{Furlanetto} S.~R.,  {Zaldarriaga} M.,   {Hernquist} L.,  2004, \mn@doi [\apj]
  {10.1086/423028}, \href
  {https://ui.adsabs.harvard.edu/abs/2004ApJ...613...16F} {613, 16}

\bibitem[\protect\citeauthoryear{{Gilfanov}, {Grimm}  \& {Sunyaev}}{{Gilfanov}
  et~al.}{2004}]{Gilfanov}
{Gilfanov} M.,  {Grimm} H.~J.,   {Sunyaev} R.,  2004, \mn@doi [\mnras]
  {10.1111/j.1365-2966.2004.07450.x}, \href
  {https://ui.adsabs.harvard.edu/abs/2004MNRAS.347L..57G} {347, L57}

\bibitem[\protect\citeauthoryear{Greig \& Mesinger}{Greig \&
  Mesinger}{2015}]{greig15}
Greig B.,  Mesinger A.,  2015, \mn@doi [Monthly Notices of the Royal
  Astronomical Society] {10.1093/mnras/stv571}, 449, 4246

\bibitem[\protect\citeauthoryear{{Grimm}, {Gilfanov}  \& {Sunyaev}}{{Grimm}
  et~al.}{2003}]{Grimm}
{Grimm} H.~J.,  {Gilfanov} M.,   {Sunyaev} R.,  2003, \mn@doi [\mnras]
  {10.1046/j.1365-8711.2003.06224.x}, \href
  {https://ui.adsabs.harvard.edu/abs/2003MNRAS.339..793G} {339, 793}

\bibitem[\protect\citeauthoryear{{G{\"u}rkan} et~al.,}{{G{\"u}rkan}
  et~al.}{2018}]{gurkan18}
{G{\"u}rkan} G.,  et~al., 2018, \mn@doi [\mnras] {10.1093/mnras/sty016}, \href
  {https://ui.adsabs.harvard.edu/abs/2018MNRAS.475.3010G} {475, 3010}

\bibitem[\protect\citeauthoryear{{Haiman}, {Rees}  \& {Loeb}}{{Haiman}
  et~al.}{1997}]{haiman97}
{Haiman} Z.,  {Rees} M.~J.,   {Loeb} A.,  1997, \mn@doi [\apj]
  {10.1086/303647}, \href
  {https://ui.adsabs.harvard.edu/abs/1997ApJ...476..458H} {476, 458}

\bibitem[\protect\citeauthoryear{Harris et~al.,}{Harris
  et~al.}{2020}]{harris2020array}
Harris C.~R.,  et~al., 2020, \mn@doi [Nature] {10.1038/s41586-020-2649-2}, 585,
  357

\bibitem[\protect\citeauthoryear{{Heesen}, {Brinks}, {Leroy}, {Heald}, {Braun},
  {Bigiel}  \& {Beck}}{{Heesen} et~al.}{2014}]{heesen14}
{Heesen} V.,  {Brinks} E.,  {Leroy} A.~K.,  {Heald} G.,  {Braun} R.,  {Bigiel}
  F.,   {Beck} R.,  2014, \mn@doi [\aj] {10.1088/0004-6256/147/5/103}, \href
  {https://ui.adsabs.harvard.edu/abs/2014AJ....147..103H} {147, 103}

\bibitem[\protect\citeauthoryear{Hunter}{Hunter}{2007}]{Hunter:2007}
Hunter J.~D.,  2007, \mn@doi [Computing in Science \& Engineering]
  {10.1109/MCSE.2007.55}, 9, 90

\bibitem[\protect\citeauthoryear{{Jana}, {Nath}  \& {Biermann}}{{Jana}
  et~al.}{2019}]{jana19}
{Jana} R.,  {Nath} B.~B.,   {Biermann} P.~L.,  2019, \mn@doi [\mnras]
  {10.1093/mnras/sty3426}, \href
  {https://ui.adsabs.harvard.edu/abs/2019MNRAS.483.5329J} {483, 5329}

\bibitem[\protect\citeauthoryear{{Lewis}, {Challinor}  \& {Lasenby}}{{Lewis}
  et~al.}{2000}]{camb}
{Lewis} A.,  {Challinor} A.,   {Lasenby} A.,  2000, \mn@doi [\apj]
  {10.1086/309179}, \href
  {https://ui.adsabs.harvard.edu/abs/2000ApJ...538..473L} {538, 473}

\bibitem[\protect\citeauthoryear{Liu, Outmezguine, Redigolo  \& Volansky}{Liu
  et~al.}{2019}]{Liu19}
Liu H.,  Outmezguine N.~J.,  Redigolo D.,   Volansky T.,  2019, \mn@doi [Phys.
  Rev. D] {10.1103/PhysRevD.100.123011}, 100, 123011

\bibitem[\protect\citeauthoryear{{Madau}, {Meiksin}  \& {Rees}}{{Madau}
  et~al.}{1997}]{madau97}
{Madau} P.,  {Meiksin} A.,   {Rees} M.~J.,  1997, \mn@doi [\apj]
  {10.1086/303549}, \href
  {https://ui.adsabs.harvard.edu/abs/1997ApJ...475..429M} {475, 429}

\bibitem[\protect\citeauthoryear{{Meiksin}}{{Meiksin}}{2021}]{Meiksin21}
{Meiksin} A.,  2021, \mn@doi [Research Notes of the American Astronomical
  Society] {10.3847/2515-5172/ac053d}, \href
  {https://ui.adsabs.harvard.edu/abs/2021RNAAS...5..126M} {5, 126}

\bibitem[\protect\citeauthoryear{{Mesinger}, {Furlanetto}  \& {Cen}}{{Mesinger}
  et~al.}{2011}]{mesinger11}
{Mesinger} A.,  {Furlanetto} S.,   {Cen} R.,  2011, \mn@doi [Monthly Notices of
  the Royal Astronomical Society] {10.1111/j.1365-2966.2010.17731.x}, \href
  {https://ui.adsabs.harvard.edu/abs/2011MNRAS.411..955M} {411, 955}

\bibitem[\protect\citeauthoryear{{Mineo}, {Gilfanov}  \& {Sunyaev}}{{Mineo}
  et~al.}{2012}]{Mineo:2012}
{Mineo} S.,  {Gilfanov} M.,   {Sunyaev} R.,  2012, \mn@doi [\mnras]
  {10.1111/j.1365-2966.2011.19862.x}, \href
  {https://ui.adsabs.harvard.edu/abs/2012MNRAS.419.2095M} {419, 2095}

\bibitem[\protect\citeauthoryear{{Mirocha} \& {Furlanetto}}{{Mirocha} \&
  {Furlanetto}}{2019}]{mirocha19}
{Mirocha} J.,  {Furlanetto} S.~R.,  2019, \mn@doi [\mnras]
  {10.1093/mnras/sty3260}, \href
  {https://ui.adsabs.harvard.edu/abs/2019MNRAS.483.1980M} {483, 1980}

\bibitem[\protect\citeauthoryear{{Mu{\~n}oz} \& {Loeb}}{{Mu{\~n}oz} \&
  {Loeb}}{2018}]{munoz18}
{Mu{\~n}oz} J.~B.,  {Loeb} A.,  2018, arXiv e-prints, \href
  {https://ui.adsabs.harvard.edu/abs/2018arXiv180210094M} {p. arXiv:1802.10094}

\bibitem[\protect\citeauthoryear{{Nusser}}{{Nusser}}{2005}]{Nusser}
{Nusser} A.,  2005, \mn@doi [\mnras] {10.1111/j.1365-2966.2005.09603.x}, \href
  {https://ui.adsabs.harvard.edu/abs/2005MNRAS.364..743N} {364, 743}

\bibitem[\protect\citeauthoryear{{Pacucci}, {Mesinger}, {Mineo}  \&
  {Ferrara}}{{Pacucci} et~al.}{2014}]{Pacucci:2014}
{Pacucci} F.,  {Mesinger} A.,  {Mineo} S.,   {Ferrara} A.,  2014, \mn@doi
  [\mnras] {10.1093/mnras/stu1240}, \href
  {https://ui.adsabs.harvard.edu/abs/2014MNRAS.443..678P} {443, 678}

\bibitem[\protect\citeauthoryear{{Planck Collaboration} et~al.,}{{Planck
  Collaboration} et~al.}{2018}]{planckcollaboration18}
{Planck Collaboration} et~al., 2018, arXiv e-prints, \href
  {https://ui.adsabs.harvard.edu/abs/2018arXiv180706209P} {p. arXiv:1807.06209}

\bibitem[\protect\citeauthoryear{{Pospelov}, {Pradler}, {Ruderman}  \&
  {Urbano}}{{Pospelov} et~al.}{2018}]{Pospelov:2018}
{Pospelov} M.,  {Pradler} J.,  {Ruderman} J.~T.,   {Urbano} A.,  2018, \mn@doi
  [\prl] {10.1103/PhysRevLett.121.031103}, \href
  {https://ui.adsabs.harvard.edu/abs/2018PhRvL.121c1103P} {121, 031103}

\bibitem[\protect\citeauthoryear{{Press} \& {Schechter}}{{Press} \&
  {Schechter}}{1974}]{press74}
{Press} W.~H.,  {Schechter} P.,  1974, \mn@doi [\apj] {10.1086/152650}, \href
  {https://ui.adsabs.harvard.edu/abs/1974ApJ...187..425P} {187, 425}

\bibitem[\protect\citeauthoryear{{Rees}}{{Rees}}{1986}]{rees86}
{Rees} M.~J.,  1986, \mn@doi [\mnras] {10.1093/mnras/222.1.27P}, \href
  {https://ui.adsabs.harvard.edu/abs/1986MNRAS.222P..27R} {222, 27P}

\bibitem[\protect\citeauthoryear{{Reis}, {Fialkov}  \& {Barkana}}{{Reis}
  et~al.}{2020}]{Reis2020}
{Reis} I.,  {Fialkov} A.,   {Barkana} R.,  2020, \mn@doi [\mnras]
  {10.1093/mnras/staa3091}, \href
  {https://ui.adsabs.harvard.edu/abs/2020MNRAS.499.5993R} {499, 5993}

\bibitem[\protect\citeauthoryear{{Reis}, {Barkana}  \& {Fialkov}}{{Reis}
  et~al.}{2022}]{reis20b}
{Reis} I.,  {Barkana} R.,   {Fialkov} A.,  2022, \mn@doi [\apj]
  {10.3847/1538-4357/ac729d}, \href
  {https://ui.adsabs.harvard.edu/abs/2022ApJ...933...51R} {933, 51}

\bibitem[\protect\citeauthoryear{{Seiffert} et~al.,}{{Seiffert}
  et~al.}{2011}]{seiffert11}
{Seiffert} M.,  et~al., 2011, \mn@doi [\apj] {10.1088/0004-637X/734/1/6}, \href
  {https://ui.adsabs.harvard.edu/abs/2011ApJ...734....6S} {734, 6}

\bibitem[\protect\citeauthoryear{{Sheth} \& {Tormen}}{{Sheth} \&
  {Tormen}}{1999}]{sheth99}
{Sheth} R.~K.,  {Tormen} G.,  1999, \mn@doi [\mnras]
  {10.1046/j.1365-8711.1999.02692.x}, \href
  {https://ui.adsabs.harvard.edu/abs/1999MNRAS.308..119S} {308, 119}

\bibitem[\protect\citeauthoryear{{Singh} et~al.,}{{Singh}
  et~al.}{2021}]{SARAS3}
{Singh} S.,  et~al., 2021, arXiv e-prints, \href
  {https://ui.adsabs.harvard.edu/abs/2021arXiv211206778S} {p. arXiv:2112.06778}

\bibitem[\protect\citeauthoryear{{Sobacchi} \& {Mesinger}}{{Sobacchi} \&
  {Mesinger}}{2013}]{sobacchi13}
{Sobacchi} E.,  {Mesinger} A.,  2013, \mn@doi [\mnras] {10.1093/mnras/stt693},
  \href {https://ui.adsabs.harvard.edu/abs/2013MNRAS.432.3340S} {432, 3340}

\bibitem[\protect\citeauthoryear{{Subrahmanyan} \& {Cowsik}}{{Subrahmanyan} \&
  {Cowsik}}{2013}]{Subrahmanyan:2013}
{Subrahmanyan} R.,  {Cowsik} R.,  2013, \mn@doi [The Astrophysical Journal]
  {10.1088/0004-637X/776/1/42}, \href
  {https://ui.adsabs.harvard.edu/abs/2013ApJ...776...42S} {776, 42}

\bibitem[\protect\citeauthoryear{{Tseliakhovich} \& {Hirata}}{{Tseliakhovich}
  \& {Hirata}}{2010}]{tseliakhovich10}
{Tseliakhovich} D.,  {Hirata} C.,  2010, \mn@doi [\prd]
  {10.1103/PhysRevD.82.083520}, \href
  {https://ui.adsabs.harvard.edu/abs/2010PhRvD..82h3520T} {82, 083520}

\bibitem[\protect\citeauthoryear{{Urry} \& {Padovani}}{{Urry} \&
  {Padovani}}{1995}]{urry95}
{Urry} C.~M.,  {Padovani} P.,  1995, \mn@doi [\pasp] {10.1086/133630}, \href
  {https://ui.adsabs.harvard.edu/abs/1995PASP..107..803U} {107, 803}

\bibitem[\protect\citeauthoryear{{Venumadhav}, {Dai}, {Kaurov}  \&
  {Zaldarriaga}}{{Venumadhav} et~al.}{2018}]{venumadhav18}
{Venumadhav} T.,  {Dai} L.,  {Kaurov} A.,   {Zaldarriaga} M.,  2018, \mn@doi
  [\prd] {10.1103/PhysRevD.98.103513}, \href
  {https://ui.adsabs.harvard.edu/abs/2018PhRvD..98j3513V} {98, 103513}

\bibitem[\protect\citeauthoryear{Virtanen et~al.,}{Virtanen
  et~al.}{2020}]{2020SciPy-NMeth}
Virtanen P.,  et~al., 2020, \mn@doi [Nature Methods]
  {10.1038/s41592-019-0686-2}, \href {https://rdcu.be/b08Wh} {17, 261}

\bibitem[\protect\citeauthoryear{{Visbal}, {Barkana}, {Fialkov},
  {Tseliakhovich}  \& {Hirata}}{{Visbal} et~al.}{2012}]{visbal12}
{Visbal} E.,  {Barkana} R.,  {Fialkov} A.,  {Tseliakhovich} D.,   {Hirata}
  C.~M.,  2012, \mn@doi [\nat] {10.1038/nature11177}, \href
  {https://ui.adsabs.harvard.edu/abs/2012Natur.487...70V} {487, 70}

\bibitem[\protect\citeauthoryear{{Zygelman}}{{Zygelman}}{2005}]{zygelman05}
{Zygelman} B.,  2005, \mn@doi [\apj] {10.1086/427682}, \href
  {https://ui.adsabs.harvard.edu/abs/2005ApJ...622.1356Z} {622, 1356}

\makeatother
\end{thebibliography}





\bsp	
\label{lastpage}
\end{document}